\documentclass{aa}  
\usepackage{natbib}
\usepackage{graphicx}
\usepackage{gensymb}
\usepackage[toc,page]{appendix}
\usepackage{pbox}
\usepackage{makecell}
\usepackage[flushleft]{threeparttable}
\usepackage{hyperref}
\usepackage{multirow}
\usepackage{multicol}
\usepackage{subcaption}
\hypersetup{colorlinks=true,breaklinks=true,citecolor=blue}
\usepackage{txfonts}
\bibpunct{(}{)}{;}{a}{}{,}

\begin{document} 

\title{X-raying winds in distant quasars: the first high-redshift wind duty cycle}

   \author{E. Bertola\inst{1}\fnmsep\inst{2}\thanks{elena.bertola2@unibo.it}
          \and
          M. Dadina\inst{2}
          \and
          M. Cappi\inst{2}
          \and
          C. Vignali\inst{1}\fnmsep\inst{2}
          \and
          G. Chartas\inst{3}
          \and
          B. De Marco\inst{4}
          \and
          G. Lanzuisi\inst{2}
          \and 
          M. Giustini\inst{5}
          \and
          E. Torresi\inst{2}
          }

   \institute{Dipartimento di Fisica e Astronomia, Universit\`a degli Studi di Bologna, via P. Gobetti 93/2, 40129 Bologna, Italy
         \and
             INAF--OAS, Osservatorio di Astrofisica e Scienza dello Spazio di Bologna, via Gobetti 93/3, 40129 Bologna, Italy
        \and 
            Department of Physics and Astronomy of the College of Charleston, Charleston, SC 29424, USA
        \and
            N. Copernicus Astronomical Center of the Polish Academy of Sciences, Bartycka 18, 00-716 Warsaw
        \and
            Centro de Astrobiolog\'ia (CSIC-INTA), Camino Bajo del Castillo s/n, Villanueva de la Ca\~nada, E-28692 Madrid, Spain
             }

   \date{Received ; accepted }
 
  \abstract
  {}
  {Theoretical models of wind-driven feedback from Active Galactic Nuclei (AGN) often identify Ultra-fast outflows (UFOs) as being the main cause for generating galaxy-size outflows, possibly the main actors in establishing the so-called AGN--galaxy co-evolution. UFOs are well characterized in local AGN but much less is known in quasars at the cosmic time when SF and AGN activity peaked ($z\simeq1$--3). It is therefore mandatory to search for evidences of UFOs in high-$z$ sources to test the wind-driven AGN feedback models. }
  {Here we present a study of Q2237+030, the Einstein Cross, a quadruply-imaged radio-quiet lensed quasar located at $z=1.695$. We performed a systematic and comprehensive temporally and spatially resolved X-ray spectral analysis of all the available \textit{Chandra} and XMM-\textit{Newton} data (as of September 2019). }
  {We find clear evidence for spectral variability, possibly due to absorption column density (or covering fraction) variability intrinsic to the source. We detect, for the first time in this quasar, a fast X-ray wind outflowing at $v_{\rm out}\simeq0.1c$ that would be powerful enough ($\dot{E}_{\rm kin}\simeq0.1 L_{\rm bol}$) to significantly affect the host galaxy evolution. We report also on the possible presence of an even faster component of the wind ($v_{\rm out}\sim0.5c$).
  Given the large sample and long time interval spanned by the analyzed X-ray data, we are able to roughly estimate, for the first time in a high-$z$ quasar, the wind duty cycle as $\approx$ 0.46\,(0.31) at 90\%\,(95\%) confidence level. Finally, we also confirm the presence of a Fe K$\alpha$ emission line with variable energy, which we discuss in the light of microlensing effects as well as considering our findings on the source. }
  {}


   \maketitle
%

\section{Introduction}
Since the discovery of the existence of scaling relations between the mass of super massive black holes (SMBH) and the global properties of their host-galaxy bulge \citep[e.g. the $ M_{\rm BH}-\sigma$ relation;][and references therein]{mcconell2011}, feedback from the active galactic nucleus (AGN) is often invoked as a key ingredient in regulating the star formation (SF) activity in the host galaxy and the growth of the SMBH itself. Despite its importance, we still lack a full comprehension of AGN feedback \citep{kormendy2013}. 

State-of-the-art models \citep[e.g.][]{kingpounds2015} predict that the SMBH/galaxy co-evolution might be established from the generation of fast accretion-disk winds, which could evolve into massive galaxy-scale outflows, possibly quenching the host galaxy star formation by sweeping out all of its interstellar medium (ISM). To provide efficient AGN feedback, the inner winds need to be accelerated at sufficient speed (nominally at semi-relativistic velocities) and need to carry a mechanical power higher than a minimum threshold set by the models at an approximate value of 0.5\%--5\%$L_{\rm bol}$ \citep[e.g.][]{dimatteo2005,hopkins2010}. In the past decade, these inner winds have often been observationally identified with the so-called Ultra-Fast Outflows \citep[UFOs;][]{tombesi2010,tombesi2011,tombesi2012,tombesi2013,gofford_2013}, which are the most extreme winds known to date, characterized by the highest outflow velocities (up to 0.2--0.3${c}$). They have been discovered in the X-ray band, through their characteristic blueshifted iron resonant absorption lines above ${\rm\approx7\ keV}$, and they are thought to arise at sub-pc scales then expand in the surrounding ISM. 

Besides probing how the inner-disk winds can trigger galaxy-scale outflows \citep{fiore2017,smith2019,mizumoto2019}, we are also still struggling to constrain the occurrence rate of UFOs. They have been mostly studied in bright and nearby Seyfert galaxies, leading to a detection fraction of $\approx$ 40--50\% \citep[][but see also \citealt{igo_2020}]{tombesi2010}. Moreover, clear variability from one observation to the other \citep{cappi2009} and also within single observations \citep{giustini2011,gofford2014} has been observed, with variations happening on time scales as short as few ks. Thus, the current idea is that UFOs might well be common and widespread, but episodic events. However, we are still rather groping in the dark for what concerns their average properties in the distant Universe. 
In particular, it is of extreme relevance to fill the gap at those redshifts where the SMBH/bulge scaling relations are thought to be shaped, i.e. where we expect the feedback processes to be most relevant and possibly visible. This corresponds to the cosmic time at which starburst and AGN activity peaked, the so-called cosmic noon ($z\simeq1$--3). 
So far, only very few sources at $z\geq1.5$ were analyzed to this goal in the X-ray band, namely APM 08279+5255, PG1115+080, H1413+117, HS 0810+2554, MG J0414+0534, \citep{hasinger2002,chartas2003,chartasH2007,chartas2016,dadina2018}. Interestingly, almost all UFO detections at $z>0.1$ are associated with gravitationally lensed quasars (GLQs), and, to our knowledge, thus far the only bright, unlensed sources at $z\geq1.5$ are PID352 and HS 1700+6416 \citep{vignali2015,lanzuisi2012}. High-redshift UFOs being associated with GLQs is not surprising given that good signal-to-noise spectra are needed to determine the presence of these winds. In this sense, the magnification provided by the gravitational lens is a unique tool to obtain good quality data in a sustainable amount of observational time. 

With this work, we performed a new and extensive study of a high-$z$ GLQ: the Einstein Cross (\object{Q2237+030}, hereafter Q2237), a quadruply-imaged quasar at $z_{\rm Q}=1.695$ (lens at $z_{\rm L}=0.039$). Discovered by Huchra in 1985 \citep{huchra1985}, Q2237 was detected for the first time in the X-ray band in 1997 by ROSAT/HRI \citep{wambsganss1999} but \textit{Chandra} was the first X-ray facility capable of resolving the four images of the quasar \citep{dai2003}. Being the first GLQ with a nearby lens to have ever been discovered, it was soon recognized as a unique case to study both macro- and microlensing properties. 
It has thus been the target of many microlensing monitoring campaigns, first in the optical, then, after the advent of \textit{Chandra}, also in the X-rays, which allowed investigating the size of the quasar's emitting regions \citep[e.g. ][]{mosquera2013,guerras2017}. Gravitational lensing theoretical models for this system agree on predicting time delays between the four source images \citep[A, B, C, D, named as in ][ -- see Fig. \ref{fig:ima_cha_431_abcd}]{yee_1988} shorter than a day ($\Delta t_{\rm AB}\approx2\ {\rm hrs}$, $\Delta t_{\rm AC}\approx-16\ {\rm hrs}$, $\Delta t_{\rm AD}\approx-5\ {\rm hrs}$) and a global magnification factor\footnote{Macro-magnification of the individual images: $\mu_{\rm A}\simeq4.6$, $\mu_{\rm B}\simeq4.5$, $\mu_{\rm C}\simeq3.8$, and $\mu_{\rm D}\simeq3.6$ \citep{schndeir1988}. } of $\mu\approx16$ \citep{schmidt1998,wertz2014}. \citet{dai2003} succeeded in confirming the shortest time delay ($\Delta t_{\rm AB}=2.7_{-0.9}^{+0.5}\ {\rm hrs}$) through the \textit{Chandra} data.
Q2237 has also been studied in the X-ray band to assess its spectral properties, either over single observations, from \textit{Chandra} \citep{chen2012} and XMM-\textit{Newton} \citep{fedorova2008}, or from \textit{Chandra} spectra stacked over multiple observations, both keeping the images separate \citep{dai2003,chen2012} and summing all the images \citep{chen2012,reynolds2014}.  
In this work, we intend to carry out the first systematic and comprehensive temporally and spatially resolved X-ray spectral analysis of this source, taking advantage of the rather complementary strengths that characterize the two X-ray facilities. In Sect. \ref{sec:data_red} we list the analyzed data and present the reduction procedures. The \textit{Chandra} and the XMM-\textit{Newton} spectra are first analyzed separately in Sects. \ref{sec:cha_sample} and \ref{sec:xmm_sample}, respectively, then all the results are combined and discussed in Sect. \ref{sec:disc_res}. Additional results obtained from the \textit{Chandra} stacked spectra are presented in Appendix \ref{app:cha}. Throughout the paper, we assume a flat ${\Lambda}$CDM cosmology \citep{planckcoll2018}, with $H_{\rm 0}=70.0\ {\rm km\ s^{-1}\ Mpc^{-1}}$ and $\Lambda_{\rm 0}=0.73$. 


\section{Data reduction}
\label{sec:data_red}
We collected, reduced and analyzed all available X-ray data of Q2237 as of September 2019: 40 archival observations in total, 37 from \textit{Chandra} and 3 from XMM-\textit{Newton} (hereafter, XMM 2002, XMM 2016 and XMM 2018), spanning over 18 years ($\sim6.7$ yr in the QSO rest frame), for a total of $\sim0.9$ Ms exposure. The Einstein Cross was the target of each pointing, so it is always observed on-axis. Tables \ref{tab:info_chandra} and \ref{tab:obs_xmm} summarize the main information of all the observations. The \textit{Chandra} observations show exposure times ranging from ${\rm7.3}$ ks to ${\rm34.2}$ ks, while those of XMM-\textit{Newton} are much longer (${\rm42.9}$ ks, ${\rm24.9}$ ks and ${\rm141.6}$ ks, in chronological order). None of the observations provides simultaneous \textit{Chandra} and XMM-\textit{Newton} data; the time elapsed between each XMM-\textit{Newton} pointing and the closest \textit{Chandra} observation ranges from one to six months. Since one of the main goals of the present work is to search for and robustly assess (through appropriate statistical tests and simulations) the presence of feedback and the significance of wind-related features in the X-ray spectra of the Einstein Cross, the lack of simultaneity between the \textit{Chandra} and XMM-\textit{Newton} data does not influence the results of this work. Conversely, this lack turns out to be quite convenient in assessing the recurrence of these features at different epochs and will allow us to investigate their presence on a more extended time baseline. 
\begin{figure}[h]
	\centering
	\includegraphics[width=1.0\linewidth]{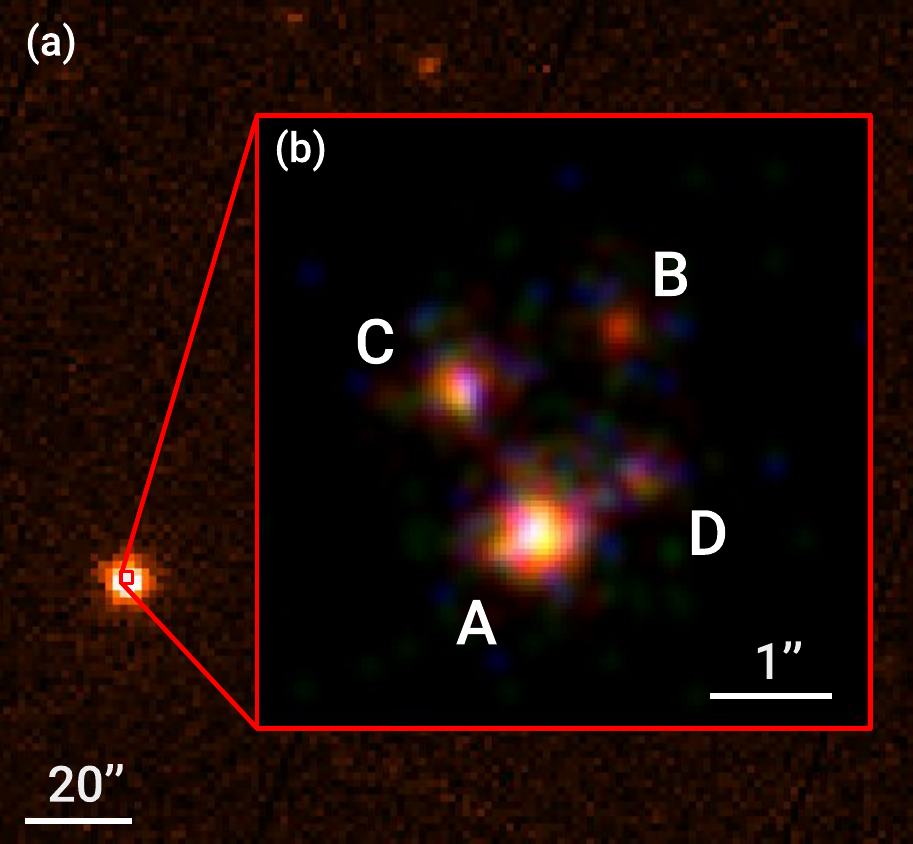} 	
	\caption{(a) EPIC-pn cleaned image of XMM 2002 in the 0.3--10 keV observed-energy band. The red square marks the $5\arcsec$ region of the \textit{Chandra} image centered on the quasar and shown in inset (b). (b) Raw \textit{Chandra} image of Q2237+030 (ObsID 431) binned with a binsize of $0.1\arcsec$, color-coded based on the observed-energy bands: 0.4--2 keV in red, 2--4.5 keV in green, 4.5--7 keV in blue. The images are named as A, B, C, and D as in \citet{yee_1988}. Given the quasar redshift ($z{\rm _Q=1.695}$), $1\arcsec$ separation corresponds to a distance of ${\rm 8.68\ kpc}$ (cosmology values: $H_{\rm 0}=70.0\ {\rm km\ s^{-1}\ Mpc^{-1}}$, $\Lambda_{\rm 0}=0.73$). }
	\label{fig:ima_cha_431_abcd}
\end{figure}

The use of data from both facilities is fundamental to our goal, for the following reasons. On the one hand, with the \textit{Chandra} data, given the satellite's superb angular resolution, we can carry out a spectral analysis that is spatially resolved over the single images of the quasar (see Fig. \ref{fig:ima_cha_431_abcd}). On the other hand, XMM-\textit{Newton} grants high counting-statistics spectra, by means of its larger effective area, that allows us to investigate the spatially integrated source emission through more complex and physical spectral models. 
All the data were reduced through the respective standard pipelines, using CIAO 4.9 and SAS 16.1, so to uniformly apply the latest calibrations to all observations. 
To extract individual-image spectra from the \textit{Chandra} data, we select the four circular regions ($r_{\rm A,B,C}=0.8\arcsec$ and $r_{\rm D}=0.6\arcsec$, with encircled energy fraction -- EEF -- at ${\rm1.5\ keV}$ of 90\% and 80\%, respectively) imposing a certain offset w.r.t. the image centroids so to limit the contamination from the neighbors. The background extraction region was selected as a source-free circle of $50\arcsec$ radius in the same chip as the target. Furthermore, to consistently analyze the data, we adopted the same regions for all the \textit{Chandra} observations, after checking that they actually corresponded to the emission peak of the individual components in each pointing. 

\begin{sidewaystable*}
\caption{Information of each \textit{Chandra} observation and details of the individual-image spectra. Those with more than 500 counts are marked with an asterisk (high-statistics sample - HSS). The observations are listed for increasing date. }
\label{tab:info_chandra}
\centering
\begin{tabular}{lcccccccccc}
\hline\hline             
ObsID &    Date    & Exposure  &   ${\rm A_{cts}}$    &   ${\rm B_{cts}}$   &	${\rm C_{cts}}$   &   ${\rm D_{cts}}$	& ${\rm A_{CR}}$ & ${\rm B_{CR}}$ & ${\rm C_{CR}}$ & ${\rm D_{CR}}$ \\ \hline
	 431  & 2000-09-06 &   30.29   & $^{\star}$1272$\pm$36 &     233$\pm$15      & $^{\star}$523$\pm$23 &	177$\pm$13	& 41.99$\pm$1.18 & 7.69$\pm$0.50  & 17.26$\pm$0.26 & 5.84$\pm$0.44  \\
	1632  & 2001-12-12 &   9.54    &      309$\pm$18      &      51$\pm$7	    &	   99$\pm$10	  &	 52$\pm$7	& 32.39$\pm$1.84 & 5.34$\pm$0.75  & 10.38$\pm$1.04 & 5.45$\pm$0.76  \\
	6831  & 2006-01-09 &   7.27    &      97$\pm$10       &      86$\pm$9	    &	   26$\pm$5	  &	 35$\pm$6	& 13.35$\pm$1.36 & 11.83$\pm$1.28 & 3.57$\pm$0.70  & 4.82$\pm$0.81  \\
	6832  & 2006-05-01 &   7.94    &      211$\pm$15      &     111$\pm$11      &	   90$\pm$10	  &	 58$\pm$8	& 26.58$\pm$1.83 & 13.98$\pm$1.33 & 11.34$\pm$1.19 & 7.31$\pm$0.96  \\
	6833  & 2006-05-27 &   7.95    &      118$\pm$11      &      56$\pm$7	    &	   54$\pm$7	  &	 21$\pm$5	& 14.83$\pm$1.37 & 7.04$\pm$0.94  & 6.79$\pm$0.92  & 2.64 $\pm$0.58 \\
	6834  & 2006-06-25 &   7.94    &      272$\pm$16      &     111$\pm$11      &	   74$\pm$9	  &    54$\pm$7   &   34.26$\pm$2.07   & 13.98$\pm$1.33	& 9.32$\pm$1.08 & 6.80$\pm$0.93  \\
	6835  & 2006-07-21 &   7.87    &      319$\pm$18      &      79$\pm$9	    &	   64$\pm$8	  &	 49$\pm$7	& 40.52$\pm$2.27 & 10.03$\pm$1.13 & 8.13$\pm$1.02  & 6.22$\pm$0.89  \\
	6836  & 2006-08-17 &   7.93    &      170$\pm$13      &      62$\pm$8	    &	   60$\pm$8	  &	 40$\pm$6	& 21.43$\pm$1.64 & 7.82$\pm$0.99  & 7.56$\pm$0.98  & 5.04$\pm$0.80  \\
	6837  & 2006-09-16 &   7.95    &      166$\pm$13      &      62$\pm$8	    &	   39$\pm$6	  &	 35$\pm$6	& 20.89$\pm$1.62 & 7.80$\pm$0.99  & 4.91$\pm$0.79  & 4.40$\pm$0.74  \\
	6838  & 2006-10-09 &   7.99    &      157$\pm$13      &      53$\pm$7	    &	   51$\pm$7	  &	 34$\pm$6	& 19.65$\pm$1.57 & 6.63$\pm$0.91  & 6.38$\pm$0.89  & 4.26$\pm$0.73  \\
	6839  & 2006-11-29 &   7.87    & $^{\star}$538$\pm$23  &     189$\pm$14      &	  108$\pm$10	  &	113$\pm$11	& 68.32$\pm$2.95 & 24.00$\pm$1.75 & 13.71$\pm$1.32 & 14.35$\pm$1.35 \\
	6840  & 2007-01-14 &   7.98    &      441$\pm$21      &     132$\pm$11      &	  118$\pm$11	  &	 84$\pm$9	& 55.29$\pm$2.63 & 16.55$\pm$1.44 & 14.79$\pm$1.36 & 10.53$\pm$1.15 \\
	11534 & 2009-12-31 &   28.46   & $^{\star}$1756$\pm$42 &     454$\pm$21      &	  164$\pm$13	  & $^{\star}$802$\pm$28 & 61.70$\pm$1.47 & 15.95$\pm$0.75 & 5.76$\pm$0.45  & 28.18$\pm$1.00 \\
	11535 & 2010-04-25 &   29.43   &      377$\pm$19      &     101$\pm$10      &	   52$\pm$7	  &	150$\pm$12	& 12.81$\pm$0.66 & 3.43$\pm$0.34  & 1.76$\pm$0.25  & 5.10$\pm$0.42  \\
	11536 & 2010-06-27 &   27.89   &      342$\pm$19      &     105$\pm$10      &	   44$\pm$7	  &	163$\pm$13	& 12.26$\pm$0.66 & 3.76$\pm$0.37  & 1.57$\pm$0.24  & 5.84$\pm$0.46  \\
	11537 & 2010-08-07 &   29.36   &      228$\pm$15      &      68$\pm$8	    &	   44$\pm$7	  &	116$\pm$11	& 7.76$\pm$0.51  & 2.31$\pm$0.28  & 1.50$\pm$0.23  & 3.95$\pm$0.37  \\
	11538 & 2010-10-02 &   29.36   & $^{\star}$501$\pm$22  &     154$\pm$12      &	   49$\pm$7	  &	423$\pm$21	& 17.06$\pm$0.76 & 5.22$\pm$0.42  & 1.67$\pm$0.24  & 14.41$\pm$0.70 \\
	11539 & 2010-11-23 &   9.83    &      93$\pm$10       &      31$\pm$6	    &	   13$\pm$4	  &	 40$\pm$6	& 9.46$\pm$0.98  & 3.15$\pm$0.57  & 1.32$\pm$0.37  & 4.09$\pm$0.64  \\
	13195 & 2010-11-26 &   9.83    &      91$\pm$10       &      23$\pm$5	    &	   10$\pm$3	  &	 27$\pm$5	& 9.25$\pm$0.97  & 2.34$\pm$0.49  & 1.01$\pm$0.32  & 2.74$\pm$0.53  \\
	13191 & 2010-11-27 &   9.83    &       82$\pm$9       &      19$\pm$4	    &	    9$\pm$3	  &	 29$\pm$5	& 8.34$\pm$0.92  & 1.93$\pm$0.44  & 0.91$\pm$0.31  & 2.95$\pm$0.55  \\
	12831 & 2011-05-14 &   29.36   & $^{\star}$2677$\pm$52 & $^{\star}$575$\pm$24 &	  215$\pm$15	  &	429$\pm$21	& 91.19$\pm$1.76 & 19.58$\pm$0.82 & 7.32$\pm$0.50  & 14.61$\pm$0.71 \\
	12382 & 2011-12-27 &   29.79   &      343$\pm$19      &     128$\pm$11      &	   59$\pm$8	  &	100$\pm$10	& 11.51$\pm$0.62 & 4.29$\pm$0.38  & 1.98$\pm$0.26  & 3.36$\pm$0.34  \\
	13960 & 2012-01-09 &   29.36   &      308$\pm$18      &     122$\pm$11      &	   52$\pm$7	  &	117$\pm$11	& 10.49$\pm$0.60 & 4.15$\pm$0.38  & 1.77$\pm$0.25  & 3.98$\pm$0.37  \\
	13961 & 2012-08-02 &   29.24   & $^{\star}$906$\pm$30  &     244$\pm$16      &	  213$\pm$15	  &	202$\pm$14	& 30.98$\pm$0.10 & 8.34$\pm$0.53  & 7.28$\pm$0.50  & 6.91$\pm$0.49  \\
	14513 & 2012-12-26 &   28.62   & $^{\star}$684$\pm$26  &     247$\pm$16      &	  339$\pm$18	  &	260$\pm$16	& 23.89$\pm$0.91 & 8.26$\pm$0.55  & 11.84$\pm$0.64 & 9.08$\pm$0.56  \\
	14514 & 2013-01-05 &   29.36   & $^{\star}$622$\pm$26  &     215$\pm$15      &	  298$\pm$17	  &	268$\pm$16	& 21.18$\pm$0.85 & 7.32$\pm$0.50  & 10.15$\pm$0.59 & 9.13$\pm$0.56  \\
	14515 & 2013-08-31 &   9.73    &      164$\pm$22      &      57$\pm$8	    &	   62$\pm$8	  &	 30$\pm$5	& 16.85$\pm$1.32 & 5.86$\pm$0.78  & 6.37$\pm$0.81  & 3.08$\pm$0.56  \\
	16316 & 2013-08-26 &   9.83    &      120$\pm$11      &      44$\pm$7	    &	   47$\pm$7	  &	 30$\pm$5	& 12.21$\pm$1.11 & 4.48$\pm$0.68  & 4.78$\pm$0.70  & 3.05$\pm$0.56  \\
	16317 & 2013-08-28 &   9.83    &      105$\pm$10      &      43$\pm$7	    &	   45$\pm$7	  &	 24$\pm$5	& 10.68$\pm$1.04 & 4.37$\pm$0.67  & 4.58$\pm$0.68  & 2.44$\pm$0.50  \\
	14516 & 2013-10-01 &   29.35   &      230$\pm$15      &     122$\pm$11      &	   70$\pm$8	  &	 96$\pm$10	& 7.83$\pm$0.52  & 4.15$\pm$0.38  & 2.38$\pm$0.29  & 3.27$\pm$0.33  \\
	14517 & 2014-05-14 &   29.36   & $^{\star}$1071$\pm$33 &     384$\pm$20      &	  322$\pm$18	  &	156$\pm$12	& 36.48$\pm$1.11 & 13.08$\pm$0.67 & 10.97$\pm$0.61 & 5.31$\pm$0.43  \\
	14518 & 2014-06-08 &   29.28   & $^{\star}$628$\pm$25  &     263$\pm$16      &	  206$\pm$14	  &	117$\pm$11	& 21.45$\pm$0.86 & 8.98$\pm$0.55  & 7.03$\pm$0.49  & 3.99$\pm$0.37  \\
	18804 & 2016-04-24 &   28.60   & $^{\star}$1009$\pm$32 &     269$\pm$16      &	  132$\pm$11	  &	126$\pm$11	& 35.28$\pm$1.11 & 9.40$\pm$0.57  & 4.61$\pm$0.40  & 4.41$\pm$0.39  \\
	19638 & 2016-12-22 &   33.34   &      302$\pm$17      &     120$\pm$11      &	   51$\pm$7	  &	 77$\pm$9	& 9.05$\pm$0.52  & 3.60$\pm$0.33  & 1.53$\pm$0.21  & 2.31$\pm$0.26  \\
	19639 & 2017-01-04 &   32.92   &      249$\pm$16      &     113$\pm$11      &	   46$\pm$7	  &	 68$\pm$8	& 7.56$\pm$0.48  & 3.43$\pm$0.32  & 1.40$\pm$0.21  & 2.06$\pm$0.25  \\
	19640 & 2017-12-20 &   34.23   &      402$\pm$20      &     210$\pm$14      &	   76$\pm$9	  &	 94$\pm$10	& 11.74$\pm$0.59 & 6.13$\pm$0.42  & 2.22$\pm$0.26  & 2.74$\pm$0.28  \\
	19641 & 2018-01-14 &   34.23   &      368$\pm$19      &     170$\pm$13      &	   92$\pm$10	  &	 77$\pm$9	& 10.75$\pm$0.56 & 4.96$\pm$0.38  & 2.10$\pm$0.25  & 2.25$\pm$0.26  \\
\end{tabular}
\tablefoot{The exposure time is given in units of ks. The observations are listed for increasing date. The image net (i.e. background-subtracted) counts (cts) and count rates (CR) are referred to the 0.4--7 keV observed-energy range. The count rates are given in units of ${\rm10^{-3}\ cts\ s^{-1}}$. }
\end{sidewaystable*}
\begin{table*}
	\caption{Information of each XMM-\textit{Newton} observation of Q2237+030. }
	\centering
	\begin{tabular}{c c c c c c c}
	    \hline\hline
		... & ObsID & Date & Exposure & Cleaned exposure & Net counts & Net count rate \\ 
		\hline
		XMM 2002 & 0110960101 & 2002-05-28 & 42.87 & 23.53 & 3837$\pm$62 & 16.3$\pm$0.3	\\
		XMM 2016 & 0781210201 & 2016-11-26 & 24.90 & 6.58 & 683$\pm$26 & 10.4$\pm$0.4	\\
		XMM 2018 & 0823730101 & 2018-05-19 & 141.60 & 108.00 & 6569$\pm$81 & 6.0$\pm$0.1	\\
		\hline
	\end{tabular}
	\label{tab:obs_xmm}
	\tablefoot{The exposure time and the cleaned exposure are given in units of ks. The source net (i.e. background-subtracted) counts and count rates are referred to the EPIC-pn spectra in the 0.3--10 keV observed-energy band. The net count rate is given in units of 10${\rm^{-2}\ cts\ s^{-1}}$. }
\end{table*}
Regarding the XMM-\textit{Newton} data, we used as extraction regions a $25\arcsec$ radius circle for the source (${\rm EEF_{Epn}\simeq85\%}$ at ${\rm1.5\ keV}$) and $80\arcsec$ radius circle for the background in all the XMM-\textit{Newton} observations. 
While the background in \textit{Chandra} is extremely low (below 0.2\% of the total counts), the first two XMM-\textit{Newton} observations were significantly affected by soft-p$^+$ flares. Operationally, we filtered the data against different count rate thresholds (CRT), then we extracted the source and background EPIC-pn spectra for each CRT and inspected how the latter relate to the former. 
The best GTI filtering threshold of the EPIC-pn data was then selected as that yielding a deviation of at least a factor of 2 between the background-subtracted source spectrum and the background spectrum in the 2--8 keV observed-energy range ($\sim$ 5.4--22 keV rest frame). 
We managed to match our criteria and obtain good quality data for XMM 2002 EPIC-pn (${\rm CRT=5.0\ cts\ s^{-1}}$, ${\rm S_{cts}=918\pm30\ cts}$ in the 2--8 keV observed-energy band). Regarding XMM 2016 EPIC-pn, mainly due to the combination of high flares and the shorter exposure of this observation, the GTI filtering threshold that satisfies our condition (${\rm CRT=2.0\ cts\ s^{-1}}$, ${\rm S_{cts}=200\pm14\ cts}$ in the 2--8 keV observed-energy band) drastically reduces the source net counts in the energy band of interest. Being XMM-\textit{Newton} data the integration over the four images of the quasar, we left this observation out of our analysis since the counting-statistics of the yielded source spectrum is so to undo the advantages provided by using XMM-\textit{Newton} data. Regarding the EPIC-MOS data, we applied the same procedure as for the EPIC-pn, using the same extraction regions (${\rm EEF_{EMOS}\simeq80\%}$ at 1.5 keV). We obtained good quality data for both cameras in XMM 2002 (${\rm CRT=1.0\ cts\ s^{-1}}$, ${\rm S_{cts}=440\pm21\ cts}$ and ${\rm S_{cts}=412\pm20\ cts}$, for MOS1 and MOS2 respectively, in the 2--8 keV observed-energy band). For completeness, we applied this procedure also on XMM 2016 EPIC-MOS 1 and 2 but, as expected given the results for the EPIC-pn spectrum and the lower effective area that EPIC-MOS 1 and 2 provide, we only managed to confirm the exclusion of this observation due to the low counting-statistic spectra obtained. XMM 2018, instead, shows a more stable background, that allowed us to select the GTI threshold directly from the detector light curves (${\rm CRT=0.9\ cts\ s^{-1}}$ and ${\rm CRT=0.2\ cts\ s^{-1}}$ for EPIC-pn and EPIC-MOS, respectively). The properties of the cleaned XMM-\textit{Newton} data are listed in Table \ref{tab:obs_xmm}. 

\subsection{Single-image multi-epoch Chandra light curves}
\label{sec:cha_lc}
\begin{figure}[!h]
	\centering
	\includegraphics[width=0.95\linewidth]{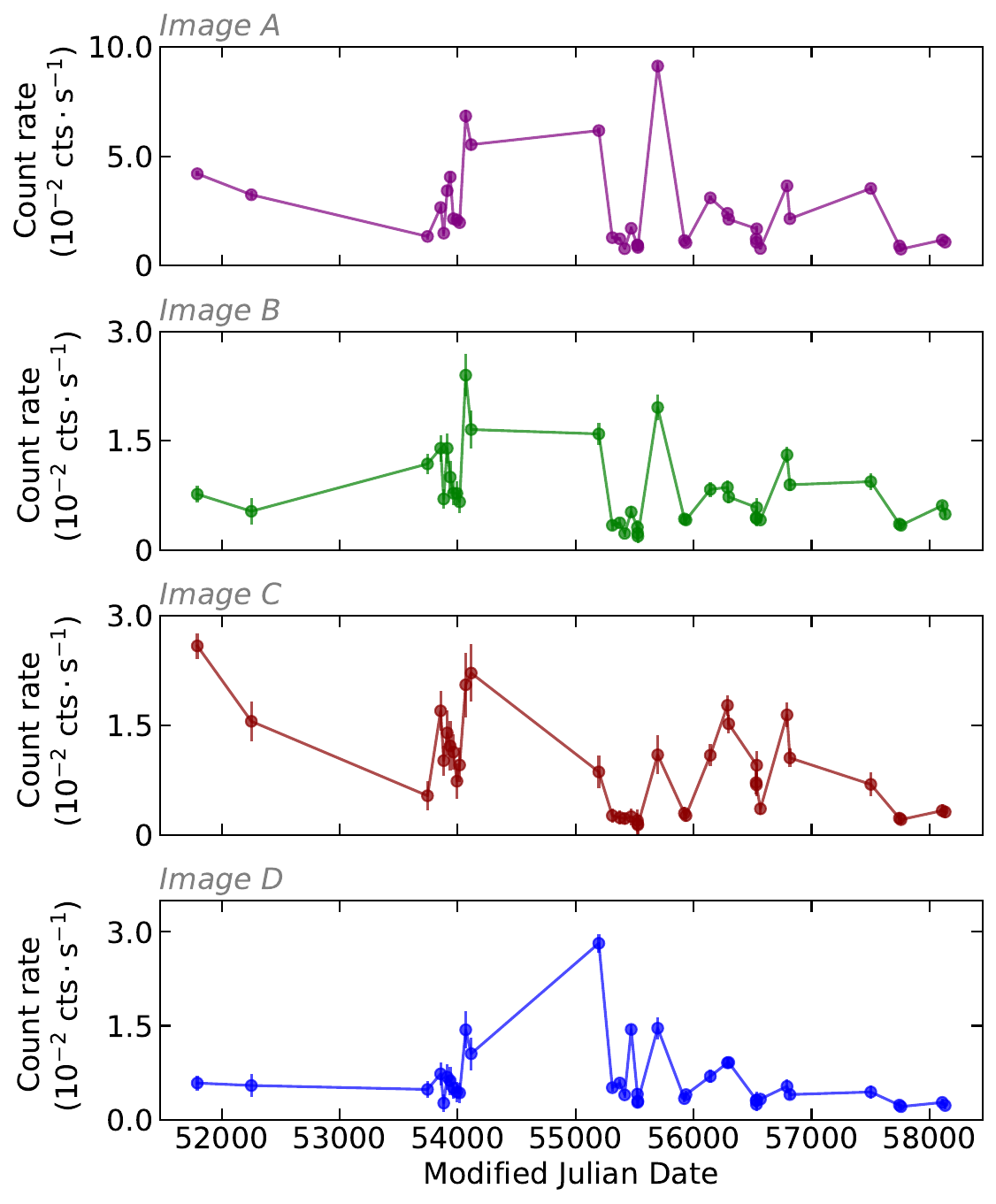}
	\caption{\textit{Chandra} individual-image multi-epoch light curves, the mean count rate between 0.4--7.0 keV observed-frame of each observation vs time. From up to bottom: image A, image B, image C, image D. The error bars are derived by the counts Poisson error. }
	\label{fig:lc_ima_cha}
\end{figure}

We produced the \textit{Chandra} single-image multi-epoch light curves by computing the image mean count rate, shown in Fig. \ref{fig:lc_ima_cha} versus the respective observation date. Each image presents flux variations up to a factor of ${\rm\approx4}$ among observations. At first glance, the mean count rates seem to vary with similar trends for all the four images. However, when taking a closer look, discrepancies between the four light curves can be seen. In fact, as expected due to the proximity of the lensing galaxy ($z_{\rm l}=0.039$) and as found by \citet{chen2011,chen2012} and \citet{dai2003} through the flux-ratio analysis, Q2237 presents microlensing events in the X-ray band that are expected to last a few months \citep{mosquera_kochanek_2011}. Moreover, \citet{dai2003} also found that more than one image of the Einstein Cross can undergo a microlensing event during a single observation. The intrinsic variability timescale of the quasar, having a $\approx1.2\times10^9\ M_{\rm\sun}$ BH mass \citep{assef2011}, is much longer than all the image time delays induced by the lens \citep{dai2003,schmidt1998}. 
\\
Therefore, the dissimilarities between the light curves in Fig. \ref{fig:lc_ima_cha} at given epoch are likely due to microlensing \citep{chen2011}. The effect of a microlensing event is to selectively magnify the emission arising from that particular portion of the background source that is behind the caustic. This results in a perturbation of the macrolensing-magnified image flux and is most relevant for the images with lower macro-magnification. An outflowing absorber moving along our line of sight (los) may produce detectable blueshifted absorption lines of highly ionized iron. We then expect microlensing events, when present, to result in a dilution of these absorption lines since they would magnify the unabsorbed emission regions that do not lie along our los. Microlensing events can then be considered as a competing effect to the detection of the UFO signatures we are mainly interested in. In this regard, they are unlikely to fake UFO absorption lines in our spectra or to shift their energy. Although a thorough analysis of the microlensing events is beyond the scope of our work, their effect on highly blueshifted absorption lines would be an interesting numerical simulation topic for a future project. 

\subsection{XMM-Newton light curves}
\label{sec:cha_xmm}
Being the BH mass of Q2237 estimated to be of the order of $10^9 M_{\rm\sun}$ \citep{assef2011}, we do not expect much short time-scale variability \citep[${\rm<1\%}$,][]{ponti2012}. 
As a sanity check, we produced the background-subtracted source light curves for the two longest observations available, XMM 2002 and XMM 2018. We grouped the light curves in 200s time bins, and split them in soft-energy (0.3--2 keV) and hard-energy (2--10 keV) bands, excluding all the background-dominated time bins (basically those at the beginning and at the end of the observation). The soft-band light curves are variable at the 95\% confidence level both for XMM 2002 and XMM 2018. The hard-band light curves are more stable, being variable at the 28\% confidence level in XMM 2002 and at 34\% in XMM 2018. We thus did not deem necessary to split the observations based on the source variability, so we extracted the time-averaged spectra integrated over the whole observations. 

\section{Chandra spectral analysis}
\label{sec:cha_sample}
We first fitted each spectrum with a single power-law modified by Galactic absorption ($N\mathrm{_H=5.1\times 10^{20}\ cm^{-2}}$; \citealt{kalberla2005}) model (Model {\sf pl}={\tt phabs*zpo}), restricting the spectral fitting to the 0.4--7 keV observed-energy range (1--19 keV rest-frame energy range). The analysis of the \textit{Chandra} spectra was then narrowed down to those with the highest counting-statistics to better constrain the presence of absorption or narrow emission/absorption features. For what concerns the lower SNR data, we analyzed their stacked spectra including all the \textit{Chandra} observations, as reported in Appendix \ref{app:cha}. 

\subsection{X-ray continuum spectral variability}
\label{sec:cha_continuum}
\begin{figure}[!h]
	\centering
	\includegraphics[width=0.95\linewidth]{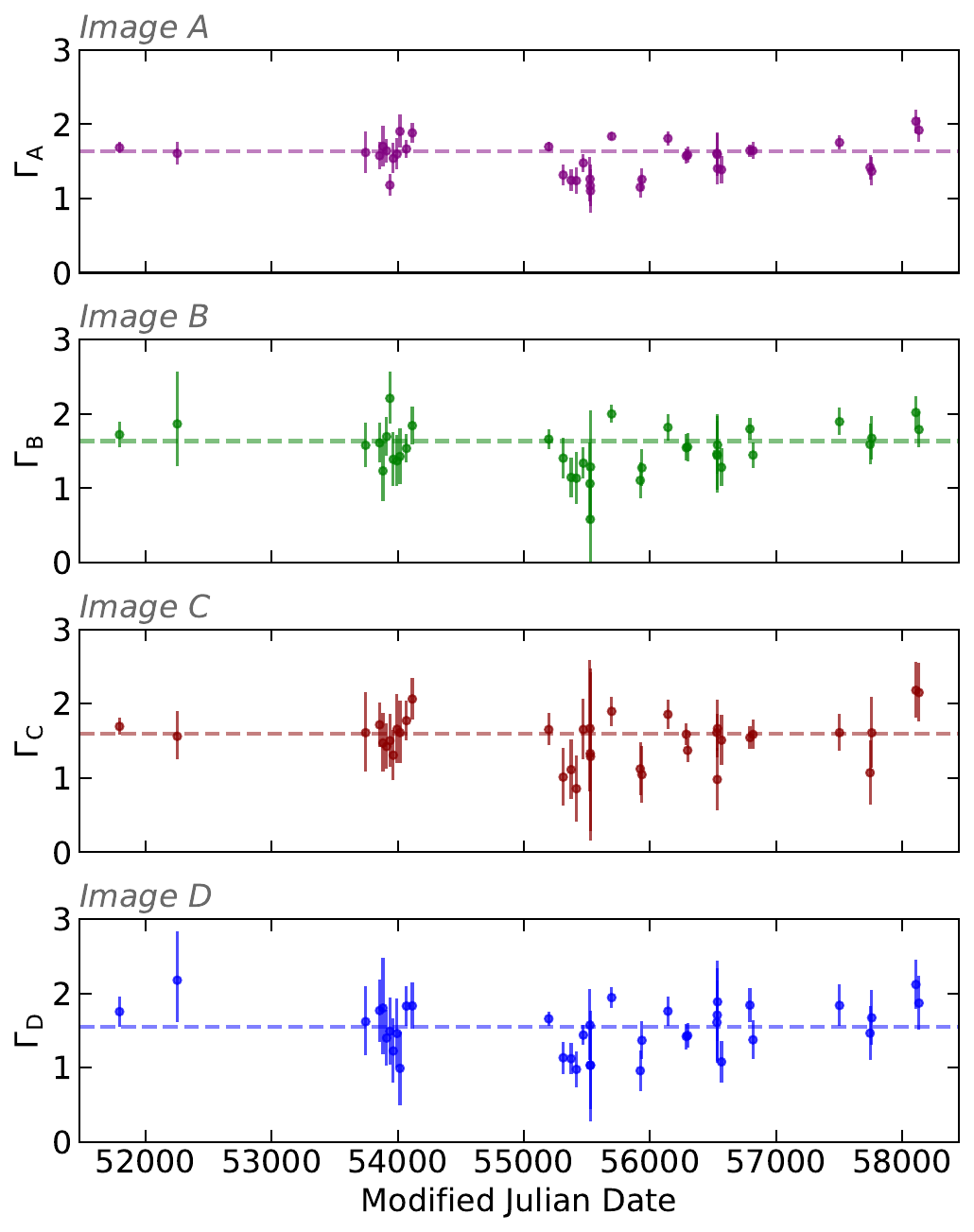}
	\caption{Variation of the photon index (${\rm1\sigma}$ errors) for each image in the \textit{Chandra} data as a function of time. The dashed line represents the best fit obtained using a constant function.} 
	\label{fig:cha_gamma}
\end{figure}

The \textit{Chandra} data allowed us to probe the source spectral variability on timescales of weeks to years. Figure \ref{fig:cha_gamma} shows the best-fit photon index ${\Gamma_i}$ ($ i{\rm=A,B,C,D}$) obtained with Model {\sf pl} as a function of time. The maximum  photon-index variation, in terms of difference between the highest and the lowest ${\Gamma_i}$ values, changes from image to image, with image A showing the smallest (${\rm\Delta\Gamma_A\approx0.94}$) and image B the largest (${\rm\Delta\Gamma_B\approx1.63}$) variations. 
By fitting with a constant (dashed lines in Fig. \ref{fig:cha_gamma}), the spectral slope was found to be significantly variable (${\rm >99.9\%}$ confidence) in each image. Considering the ratios of the $\Gamma_i$, we can check whether the photon-index variations are intrinsic (if common to all images) or induced by microlensing. We find the ratios consistent with being overall constant and approximately equal to one. This agrees with the approximation applied by \citet{chen2012}, who analyzed the first 20 \textit{Chandra} observations of our list linking the photon index of the four images at any given epoch. Moreover, the maximum $\Gamma$ variations of all the images are consistent with each other when propagating their 1$\sigma$ errors, thus no image shows a significantly higher maximum photon-index variation w.r.t. the others. Therefore, we can assume the variations of the continuum to be overall coherent between the four images, i.e. not induced by microlensing but inherent to the quasar. 

\begin{figure*}[!h]
	\centering
	\includegraphics[width=1\linewidth,keepaspectratio]{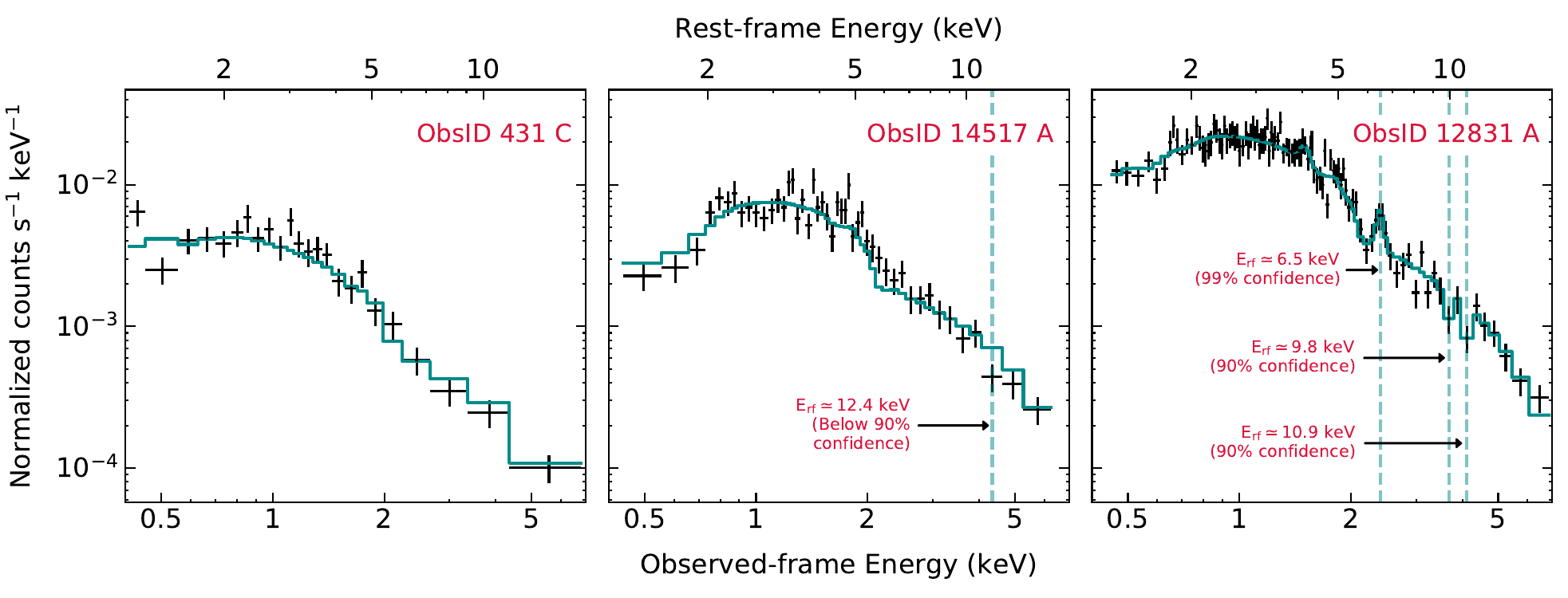}
	\caption{Data (black) and best-fit model (blue) for three of the \textit{Chandra} HSS spectra, representative of three statistics regimes. From left to right: ObsID 431 C (least counts), 14517 A (intermediate counts), and 12831 A (highest counts). The dashed vertical lines indicate the energies of the emission/absorption lines found adopting the blind-search method of \citet{tombesi2010}. Only those above 90\% were included in the best-fit models (i.e. those in Tables \ref{tab:cha_high_lines_em} and \ref{tab:cha_high_lines_abs}).}
	\label{fig:cha_spectra}
\end{figure*}

To investigate the presence of any intrinsic spectral variability, we restrict our analysis to the high-statistics sample (HSS) to better constrain the best-fit spectral parameters. This sub-sample is made of the fourteen spectra that show more than 500 source net counts in the 0.4--7 observed-energy range (those marked by a star in Table \ref{tab:info_chandra}). The count threshold was selected to allow us to apply the $\chi^2$ statistics after binning the source spectra to at least 20 cts/bin. Fourteen spectra were extracted from eleven observations, since in three epochs (ObsIDs 431, 11534, 12831) two images exceed our threshold. In Fig. \ref{fig:cha_spectra} we show three representative HSS spectra: that with the most counts (ObsID 12831 A), one of those with the least (ObsID 431 C) and one with an intermediate number of counts (ObsID 14517). Spectra with lower-counting statistics were used to produce stacked spectra, both keeping the images separate and combining the four images, the results of which are presented in Appendix \ref{app:single_ima_stacked} and Appendix \ref{app:all_combined}, respectively. We anticipate that the results obtained with the stacked spectra are overall consistent with the ones presented here. 
\renewcommand{\arraystretch}{1.15}
\begin{table}[h]
	\centering
	\caption{Summary of the best-fit parameters for Model {\sf pl\_a} ({\tt phabs*zphabs*zpowerlw}) when applied to the high-statistics sample. Those that actually require extra absorption at a significance level above 99\% according to the F-test are in boldface. } 
	\begin{tabular}[!h]{c c c c c}
		\hline\hline
		ObsID      & $\Gamma$                 & $N{\rm_H}$             & ${\rm\Delta \chi^2}$ & Confidence         \\ \hline
		431     A  & $1.86_{-0.12}^{ +0.12}$  & $0.34_{-0.20}^{ +0.20}$ & 8.4                  & 98.8\%             \\
		431		 C & $1.90_{-0.36}^{ +0.42}$  & $0.53_{-0.20}^{ +0.23}$ & 5.9                  & 97.7\%             \\
		6839     A & $1.80_{-0.20}^{ +0.20}$  & $0.46_{-0.42}^{ +0.52}$ & 3.2                  & 84.6\%             \\
		11534    A & $1.91_{- 0.10}^{ +0.11}$ & $0.62_{-0.25}^{ +0.29}$ & 19.3                 & \textbf{> 99.99\%} \\
		11534	 D & $2.01_{ -0.18}^{ +0.20}$ & $1.28_{-0.48}^{ +0.57}$ & 24.3                 & \textbf{> 99.99\%} \\
		11538    A & $2.06_{- 0.26}^{ +0.28}$ & $2.88_{-1.17}^{ +1.40}$ & 22.0                 & \textbf{99.9\%}    \\
		12831    A & $1.98_{ -0.09}^{ +0.09}$ & $0.27_{-0.21}^{ +0.22}$ & 4.7                  & 96.7\%             \\
		12831	 B & $2.10_{ -0.24}^{ +0.27}$ & $0.50_{-0.49}^{ +0.60}$ & 2.9                  & 95.6\%             \\
		13961    A & $2.02_{ -0.16}^{ +0.18}$ & $0.67_{-0.44}^{ +0.53}$ & 7.0                  & 97.7\%             \\
		14513    A & $1.80_{ -0.18}^{ +0.20}$ & $0.88_{-0.56}^{ +0.67}$ & 7.1                  & 98.4\%             \\
		14514    A & $1.79_{ -0.20}^{ +0.22}$ & $0.77_{-0.66}^{ +0.80}$ & 3.7                  & 85.6\%             \\
		14517    A & $1.79_{ -0.13}^{ +0.14}$ & $0.59_{-0.40}^{ +0.49}$ & 6.5                  & 97.2\%             \\
		14518    A & $1.84_{ -0.22}^{ +0.24}$ & $1.01_{-0.84}^{ +1.03}$ & 4.1                  & 96.5\%             \\
		18804    A & $2.10_{ -0.18}^{ +0.19}$ & $1.45_{-0.71}^{ +0.85}$ & 13.8                 & \textbf{99.9\%}    \\ \hline\hline
	\end{tabular}
	\label{tab:cha_high_zphabs}
	\tablefoot{Col. 1: ObsID and image; Col. 2: photon index; Col. 3: column density in excess to the Galactic value (in units of ${\rm 10^{22}\ cm^{-2}}$ and placed at $z=1.695$); Col. 4: ${\rm\Delta \chi^2\ ^b}$ w.r.t. Model {\sf pl} (${ \Delta dof=1}$); Col. 5: F-test confidence level. All the errors are computed at 90\% confidence level for one parameter of interest.}
\end{table}
\renewcommand{\arraystretch}{1.0}

\begin{figure}[h]
	\centering
	\includegraphics[width=1.0\linewidth]{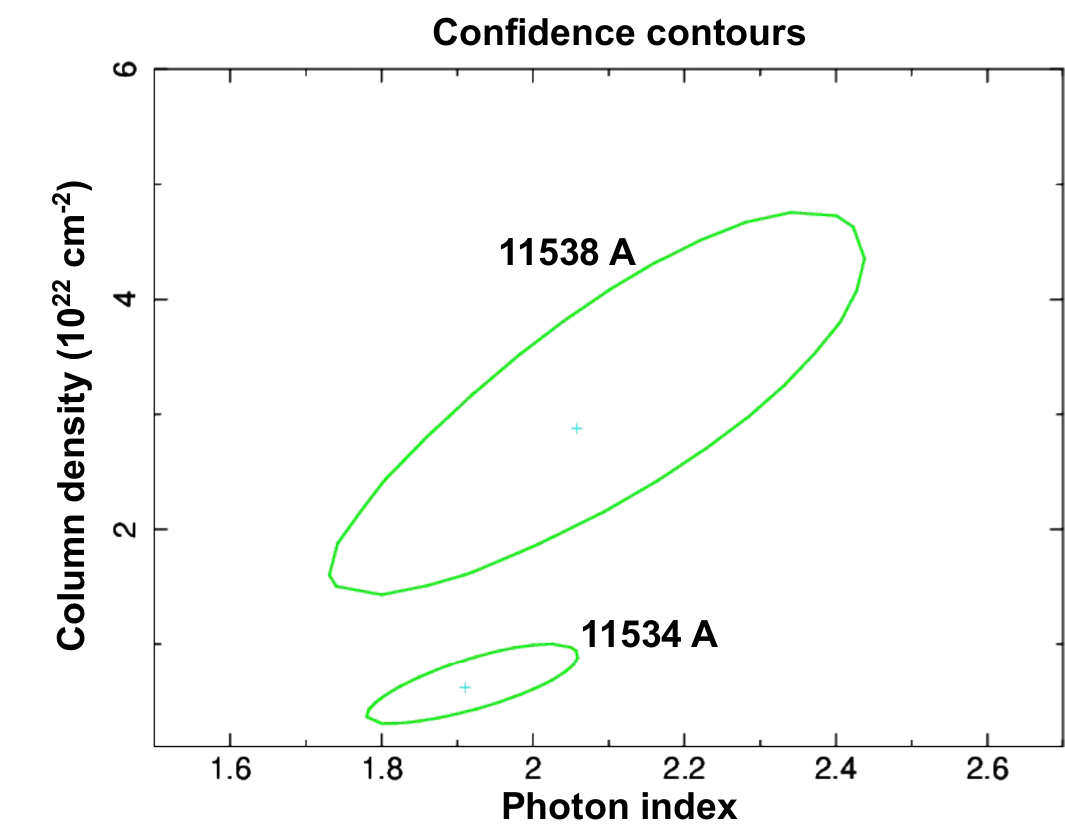}
	\caption{90\% confidence contours of $N_{\rm H}$ vs. $\Gamma$ for ObsIDs 11534 A and 11538 A, which correspond to the spectra that show the largest difference in column density, among the four that require a cold absorber at more than 99\% confidence level. }
	\label{fig:cha_cont_nH_superpos}
\end{figure}

To search for additional spectral continuum complexities for the HSS, we first modified Model {\sf pl} by adding a {\tt zphabs} component, accounting for photo-electric absorption of the primary emission in a cold medium (Model {\sf pl\_a}={\tt phabs*zphabs*zpo}), which we placed at the quasar's redshift. We find that only four spectra\footnote{ObsIDs 11534 A and D, 11538 A, 18804 A. } of the HSS require extra absorption at more than 99\% confidence level (evaluated through the F-test), while for the other ten we could only derive an upper limit to such additional column density. 
All the best-fit values and the respective F-test significance are listed in Table \ref{tab:cha_high_zphabs}, with those at more than 99\% confidence level shown in boldface. Among the four spectra highly requiring some extra absorption, two are referred to two images from the same epoch (ObsID 11534 image A and D) and show consistent column densities within 1$\sigma$ errors. Considering only the cases where extra absorption is required by the data, we find the column density to be variable at more than 99\% confidence throughout the three epochs. 
To test the assumption on the location of the absorber, we compared the results obtained for the two spectra (ObsID 11534 A and 11538 A) showing the largest variation in column-density values by plotting their 90\% confidence contours of $N_{\rm H}$ as a function of the photon index. As shown in Fig. \ref{fig:cha_cont_nH_superpos}, their column densities are not consistent, while their photon indices are. Furthermore, the time interval among the two observations (see Table \ref{tab:info_chandra}) is ${\rm \sim275\ d}$ in the observer frame, which means ${\rm \sim102\ d\simeq0.3\ yrs}$ in the quasar rest frame. Considering that the two spectra are referred to the same image, we interpret the $N_{\rm H}$ discrepancy and the short time elapsed as indications of the extra absorption being dominated by the component at the redshift of the quasar. 

\subsection{Narrow emission/absorption features}
\label{sec:cha_lines}
We next searched for emission/absorption features, again only in the HSS spectra to obtain better constrains.

A first blind search is carried out by applying the method developed by \citet{tombesi2010}. By stepping both the energy and the normalization of a Gaussian component, it allows us to visualize, as a function of both parameters, the statistical improvement produced by the addition of a narrow feature (${\rm\sigma=0.01\ keV}$) in terms of ${\rm\Delta\chi^2}$ translated in confidence contours for the addition of two parameters.
We then build the best-fit models of all the HSS spectra by adding a Gaussian component for each emission/absorption line indicated at more than 90\% confidence by the blind search. Even though it is known not to be reliable when assessing the significance of narrow features \citep{protassov2002}, we compute the F-test significance for each line to have a slightly better constrain than that of the blind search and only keep those that are still above 90\% confidence level. Finally, we evaluate the actual significance of the absorption lines at $E{\rm _{rf}>7\ keV}$ by building a Bayesian posterior predictive probability distribution through Monte Carlo simulations, as argued by \citet{protassov2002}.

We applied the \citet{tombesi2010} tool over the 0.4--5.0 keV observed energy range ($\sim$1--13 keV rest-frame energy band), that corresponds to the range of interest for both soft X-ray features and iron resonant lines. The upper energy limit was set to exclude the bins with the least SNR. We selected as baseline models either Model {\sf pl} or Model {\sf pl\_a} based on the requirement of extra absorption (see Table \ref{tab:cha_high_zphabs}).  
\renewcommand{\arraystretch}{1.15}
\begin{table}[h]
	\centering
	\caption{Rest-frame energies and equivalent widths of (a) the emission and (b) the absorption lines detected at more than 90\% confidence in the high-statistics sample, based on the F-test. Those showing more than 99\% confidence are reported in boldface. }
	\label{tab:cha_high_lines}
	\subcaption{Emission lines}
		\begin{tabular}[!h]{ccccl}
			\hline\hline
			ObsID      & $E{\rm_{line}}$          & EW                & Cont.         & F-test      \\ \hline
			431      A & $5.90_{- 0.36}^{ +0.31}$  & $368_{-278}^{ +277}$  & 90\%          & 90.4\%         \\
			11534    A & $3.65_{- 0.10}^{ +0.13}$  & $101_{-68}^{ +69}$  & 90\%          & 93\%           \\
			           & $5.35_{- 0.12}^{ +0.13}$  & $179_{-111}^{ +111}$  & 90\%          & 93\%           \\
			11534    D & $2.25_{- 0.10}^{ +0.10}$  & $163_{-107}^{ +137}$  & 90\%          & 98.5\%         \\
			12831    A & $4.04_{ -0.15}^{ +0.14}$  & $75_{-57}^{ +57}$  & 90\%          & 93\%           \\
			           & $6.47_{ -0.12}^{ +0.11}$  & $284_{-148}^{ +148}$  & 99\% & \textbf{99.7\%}         \\
			13961    A & $3.78_{ -0.12}^{ +0.16}$  & $173_{-105}^{ +105}$  & 90\%          & 96\%           \\ \hline
		\end{tabular}
	\label{tab:cha_high_lines_em}
	\bigskip
	\subcaption{Absorption lines}
		\begin{tabular}[!h]{cccclc}
			\hline\hline
			ObsID      & $E{\rm_{line}}$          & EW                 & Cont.         & F-test & MC     \\ \hline
			431      A & $11.92_{- 1.69}^{ +0.23}$ & $1181_{-575}^{ +778}$ & 99\% & 99.5\% &\textbf{ 99.5\% }\\
			11534    A & $8.01_{- 0.14}^{ +0.42}$  & $287_{-185}^{ +212}$ & 90\%          & 97\%   & 90.3\% \\
			12831    A & $9.78_{- 0.18}^{ +0.43}$  & $255_{-169}^{ +211}$ & 90\%          & 95\%   & 84.2\% \\
			& $10.90_{- 0.20}^{ +0.53}$ & $368_{-196}^{+258}$  & 99\% & 99.1\% & 94.9\% \\
			13961    A & $9.39_{ -0.08}^{ +1.90}$  & $1025_{-532}^{+542}$ & 99\% & 98.3\% & \textbf{99.1\%} \\
			14514    A & $12.51_{-2.28}^{ +0.10}$  & $1861_{-1273}^{+1354}$ & 90\%          & 92\%   & 98.4\% \\
		\hline
		\end{tabular}
		\label{tab:cha_high_lines_abs}
		\tablefoot{Col. 1: ObsID and image; Col. 2: line energy (in units of keV); Col. 3: line rest-frame equivalent width (in units of eV); Col. 4: blind-search confidence level; Col. 5: F-test confidence level; Col. 6: Monte-Carlo-simulation confidence level. All the errors are computed at 90\% confidence level for one parameter of interest. The energy width of the lines is set to 0.01 keV. }
\end{table}
\renewcommand{\arraystretch}{1.0}
We then obtain the best-fit models by adding a narrow {\tt zgauss} component for each line indicated at a confidence above 90\% from the blind search. We only keep those lines having a significance above 90\% confidence both from the blind search and the F-test, which are summarized in Table \ref{tab:cha_high_lines}. This procedure indicated the presence (at 90\% confidence level) of blueshifted iron resonant absorption lines in five spectra out of fourteen. 

Finally, we evaluate through Monte Carlo simulations the actual significance of the absorption lines in Table \ref{tab:cha_high_lines_abs}. Following \citet{protassov2002}, each of these five spectra was simulated 1000 times through the {\tt XSPEC} {\tt fakeit} function from the respective null model (Model {\sf pl} or Model {\sf pl\_a} if extra absorption required -- see Table \ref{tab:cha_high_zphabs}). 
This confirmed that all the absorption lines at $E_{\rm rf}>7\ {\rm keV}$ are detected above 90\% confidence but the one at 9.8 keV in ObsID 12831 A (see Table \ref{tab:cha_high_lines_abs}). Thus, we find blueshifted iron resonant absorption lines in five epochs out of the eleven included in the HSS. 
Through the binomial distribution, we then evaluate the global probability of detecting these absorption lines in five spectra out of a sample of fourteen by chance. We conservatively considered all the lines as detected at ${\rm 90\%}$ confidence, even though more than half show higher significance. The probability of a by-chance detection is $P=7.76\cdot10^{-3}$, yielding an overall significance of 99.2\% (i.e. slightly below $3\sigma$) for the detection of these absorption lines at $E_{\rm rf}>7\ {\rm keV}$ throughout the HSS. 

\begin{figure}[!h]
	\includegraphics[width=1.0\linewidth]{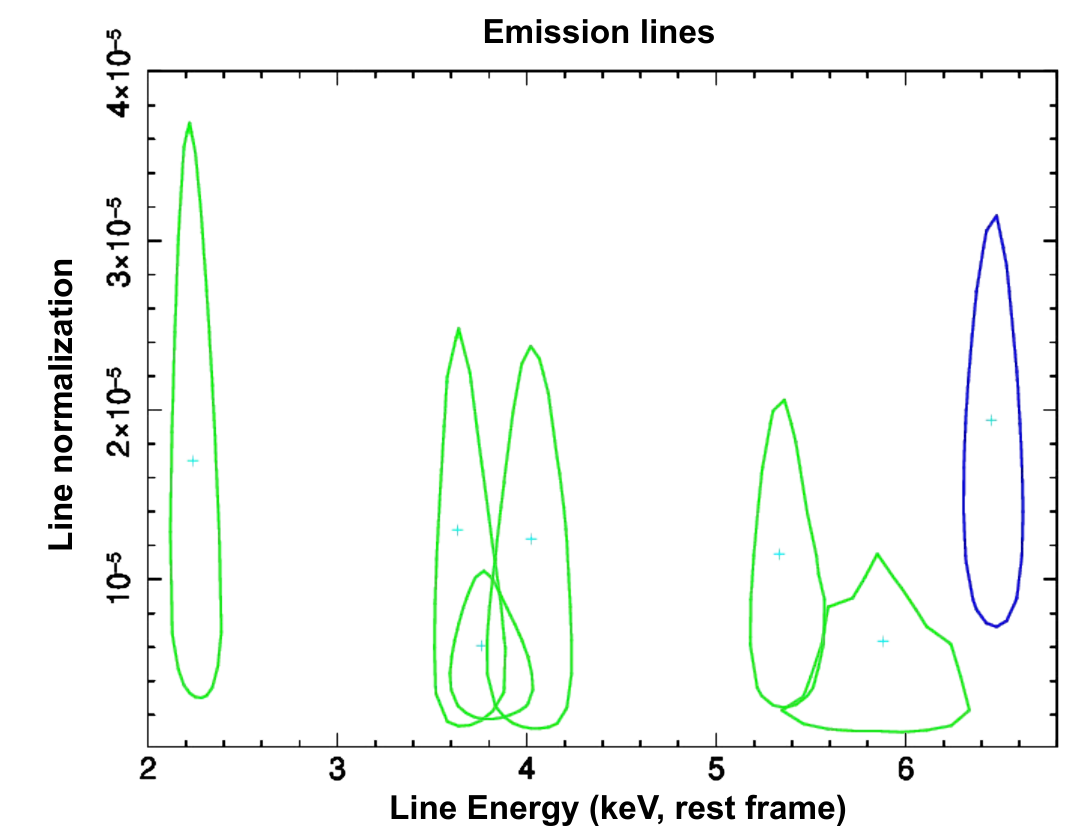}
	\caption{90\% energy-normalization confidence contours (1.6$\sigma$) for the emission lines reported in Table \ref{tab:cha_high_lines_em} (ObsIDs 431 A, 11534 A, 11534 D, 12831 A, 13961 A). The blue contour corresponds to the line detected at more than 99\% confidence level in ObsID 12831 A (based on the F-test significance). }
	\label{fig:cha_high_superpos_line_em}
\end{figure}

\begin{figure}[h]
	\includegraphics[width=1.0\linewidth]{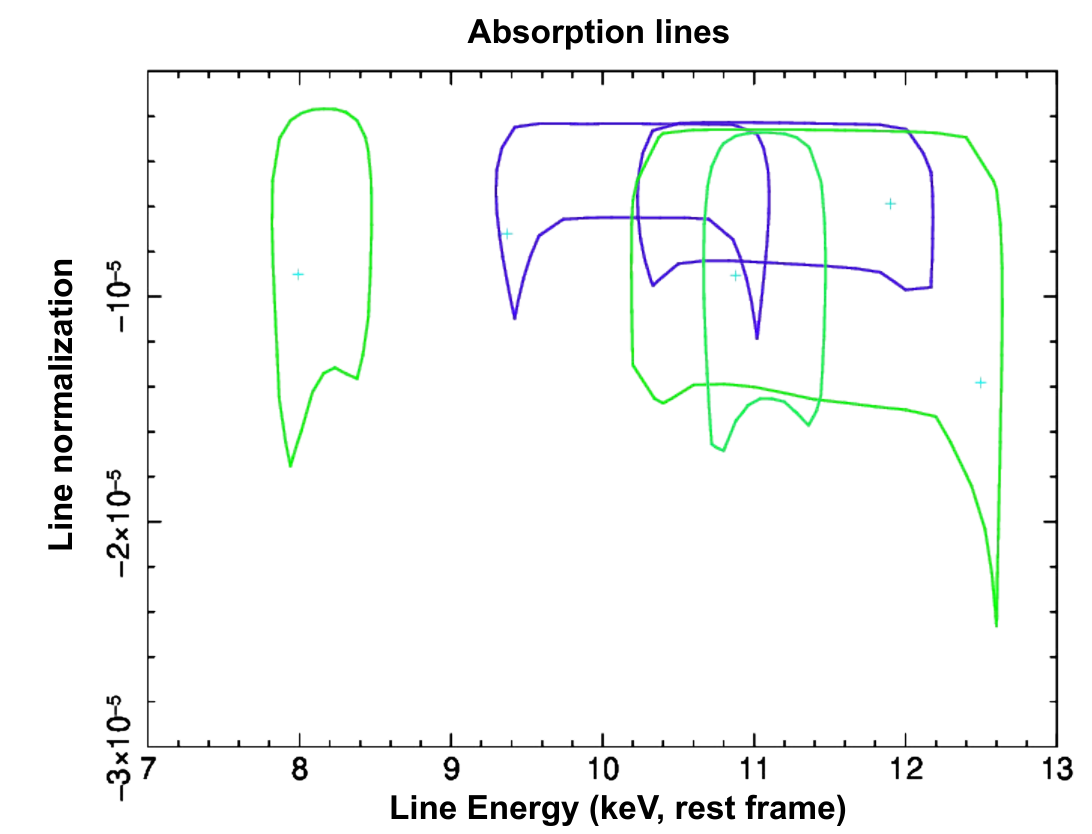}
	\caption{90\% energy-normalization confidence contours (1.6$\sigma$) for the absorption lines reported in Table \ref{tab:cha_high_lines_abs} (ObsIDs 431 A, 11534 A, 12831 A, 13961 A, 14514 A). Those in blue correspond to the lines detected at more than 99\% confidence level (based on the Monte-Carlo-simulation significance). }
	\label{fig:cha_high_superpos_line_abs}
\end{figure}

To inspect the persistence of such features through the different epochs, we overlapped the $1.6\sigma$ confidence contours of the narrow emission and absorption lines separately (Figs. \ref{fig:cha_high_superpos_line_em} and \ref{fig:cha_high_superpos_line_abs}, respectively); the 90\% confidence contours of the features detected at more than 99\% confidence are reported in blue, those of the other lines are in green.  

The emission lines (Fig. \ref{fig:cha_high_superpos_line_em}) span over the 2.2--6.5 keV rest-frame energy band. The microlensed Fe K$\alpha$ line found by \citet{dai2003} in the combined spectra of ObsIDs 431 A and 1632 A ($E=5.7_{-0.3}^{+0.2}\ {\rm keV}$, ${\rm\sigma=0.87_{-0.15}^{-0.30}\ keV}$) is only marginally detected (90\% confidence) in the spectrum of ObsID 431 A as a narrow line, probably due to the fact that we are analyzing single-epoch spectra while \citet{dai2003} stacked the first two observations. The energy of the highly significant emission line in ObsID 12831 A ($E_{\rm rf}= 6.47_{-0.12}^{+0.11}\ {\rm keV}$) is consistent, within ${\rm1.6\ \sigma}$ errors, with the centroid energy of the skewed line found by \citet{reynolds2014} ($E= 6.58\pm0.03\ {\rm keV}$) in the combined spectra of all the images, stacking the first 26 observations (ObsIDs 431 -- 14514). Following \citet{dai2003} and \citet{chartas2016,chartas2017}, the remaining emission lines can be associated with microlensed Fe K$\alpha$ lines. 

The absorption features (Fig. \ref{fig:cha_high_superpos_line_abs}) show a group of values around ${\rm11\ keV}$ and one line at about 8 keV. Each of the confidence contours of the lines clustered around ${\rm11\ keV}$ cover a wide range of energy. This could be interpreted as being due to the blend of two or more lines given the lower SNR at the higher energies, thus larger energy bins, or ascribed to intrinsic very broad lines as seen in other high-$z$ lensed and unlensed quasars, for instance APM 08279+5255 \citep{chartas2009} and HS 1700+6416 \citep{lanzuisi2012}. By letting the width of the absorption lines in Table \ref{tab:cha_high_lines_abs} free to vary, we find them to be consistent with zero, with upper limits ranging from ${\rm0.8\ keV}$ (ObsID 12831 A) up to ${\rm3.2\ keV}$ (ObsID 13961). The feature in ObsID 13961 A is the most peculiar; it would actually prefer an intrinsically broad (${\rm\sigma=2.2_{-0.5}^{+1.0}\ keV}$) line but the low statistics prevent us to simultaneously constrain the three line parameters (energy, width, normalization). As a result, we consider all the lines in Table \ref{tab:cha_high_lines_abs} consistent with being narrow. 

The highly significant absorption lines (blue contours in Fig. \ref{fig:cha_high_superpos_line_abs}) are both found at energies above ${\rm9\ keV}$ (rest frame) and are consistent with each other at 1.6$\sigma$. These features are detected in observations 431 and 13961, which are separated by twelve years (${\rm\sim4.5}$ yrs proper time), and are also consistent in energy with the slightly less significant features of ObsIDs 12831 and 14514. Thus, we infer that the lines at $E_{\rm rf}>9\ {\rm keV}$ are presumably produced by the same highly ionized outflow, which, from the period elapsed between the observations, is most likely variable. Interestingly, there seems to be no clear correlation between the additional absorption found in Sect. \ref{sec:cha_continuum} and the detection of these absorption lines above ${\rm7\ keV}$. In fact, only one spectrum (ObsIDs 11534 A) out of the four requiring an additional cold absorber present such lines above 90\% confidence. 

Finally, we analyzed the combined spectra of images B+C+D from ObsIDs 431, 12831 and 13961, i.e. those observations in which image A shows highly significant absorption and/or emission lines (see Tables \ref{tab:cha_high_lines_em} and \ref{tab:cha_high_lines_abs}). In absence of microlensing events, since the time delays between the images are short when compared to the intrinsic variability timescale, one would expect the B+C+D combined spectra to show the same kind of features as the image A spectra at a confidence at least higher than 90\%, since the number of counts of the stacked spectra is similar to that of the respective image A. We find that the B+C+D spectra do not present any of the lines of Tables \ref{tab:cha_high_lines_em} and \ref{tab:cha_high_lines_abs} at more than 90\% confidence. However, the upper limits we derive on their equivalent widths (EWs) are consistent with those of the respective image A lines. Our interpretation is that one, or more than one, of images B, C, and D is microlensed, thus the absorption lines are smeared out in the individual image spectra and the Fe K$\alpha$ emission lines are likely shifted to higher/lower energies, getting even more diluted when we stack the images together. 

\subsection{Summary of the Chandra spectral results}
\begin{itemize}
    \item We find clear spectral index variability at a significance larger than 99\% as inferred from the analysis of the whole \textit{Chandra} dataset (Fig. \ref{fig:cha_gamma}). From the $\Gamma$ ratios, those variations seem to be intrinsic and not induced by microlensing. Photon-index variability might also be induced by variable absorption in some observations, as the HSS analysis pointed out (Table \ref{tab:cha_high_zphabs}). Having a variability timescale of ${\rm\sim0.3\ yrs}$ (rest frame), such absorber is most likely dominated by an in situ component. 
    \item Five spectra of the HSS do also show narrow emission lines below ${\rm 3.5\ keV}$ in the observed frame (significance above ${\rm90\%}$ confidence - Table \ref{tab:cha_high_lines_em}). In one spectrum (ObsID 12831 A) we detect a highly significant narrow emission line at $E_{\rm rf}=6.47_{-0.12}^{+0.11}\ {\rm keV}$; this energy is consistent with that detected by \citet{reynolds2014}. The line found by \citet{dai2003} is only marginally detected in the spectrum of ObsID 431 A.
    \item Finally, five spectra show narrow absorption features in the 3--5 keV observed-frame energy band ($\sim$ 8--13.5 keV rest frame) at more than 90\% confidence (assessed through Monte Carlo simulations - Table \ref{tab:cha_high_lines_abs}). A certain recurrence and/or persistence of those features in this energy range is indicated by the consistency of their confidence contours (Fig. \ref{fig:cha_high_superpos_line_abs}), although we note that the error associated with the energy of the lines is typically large. The overall significance of detecting such features in the HSS is proved to be almost 3$\sigma$. 
\end{itemize}

\section{XMM-Newton spectral analysis}
\label{sec:xmm_sample}
Having assessed the source spectral variability and the presence of spectral complexities through the \textit{Chandra} observations, we then analyzed the XMM-\textit{Newton} spectra, and attempted to model those complexities with reflection and complex absorption models. Both XMM 2002 and XMM 2018 spectra were grouped in order to obtain at least 20 counts per bin. Given the large number of counts (Table \ref{tab:obs_xmm}), we set the minimum energy width of each bin at one third of the CCD resolution using the task {\tt specgroup} within SAS, so not to oversample the energy resolution of the instrument. Due to background-dominated bins at the higher energies, the spectral fitting of XMM 2002 and XMM 2018 data was performed in the 0.3--8 keV observed-energy range (0.8--22 keV rest-frame energy range) and in the 0.3--7.0 keV observed-energy range (1--19 keV rest-frame energy range), respectively, so to obtain more reliable results. For both observations, we analyze and report here only the EPIC-pn spectra which deliver the best SNR but our results were also compared, and confirmed, by checking their consistency with the EPIC-MOS data. 

\subsection{XMM 2002}
\label{sec:xmm_2002}
\begin{figure}[h]
	\centering
	\includegraphics[width=0.97\linewidth]{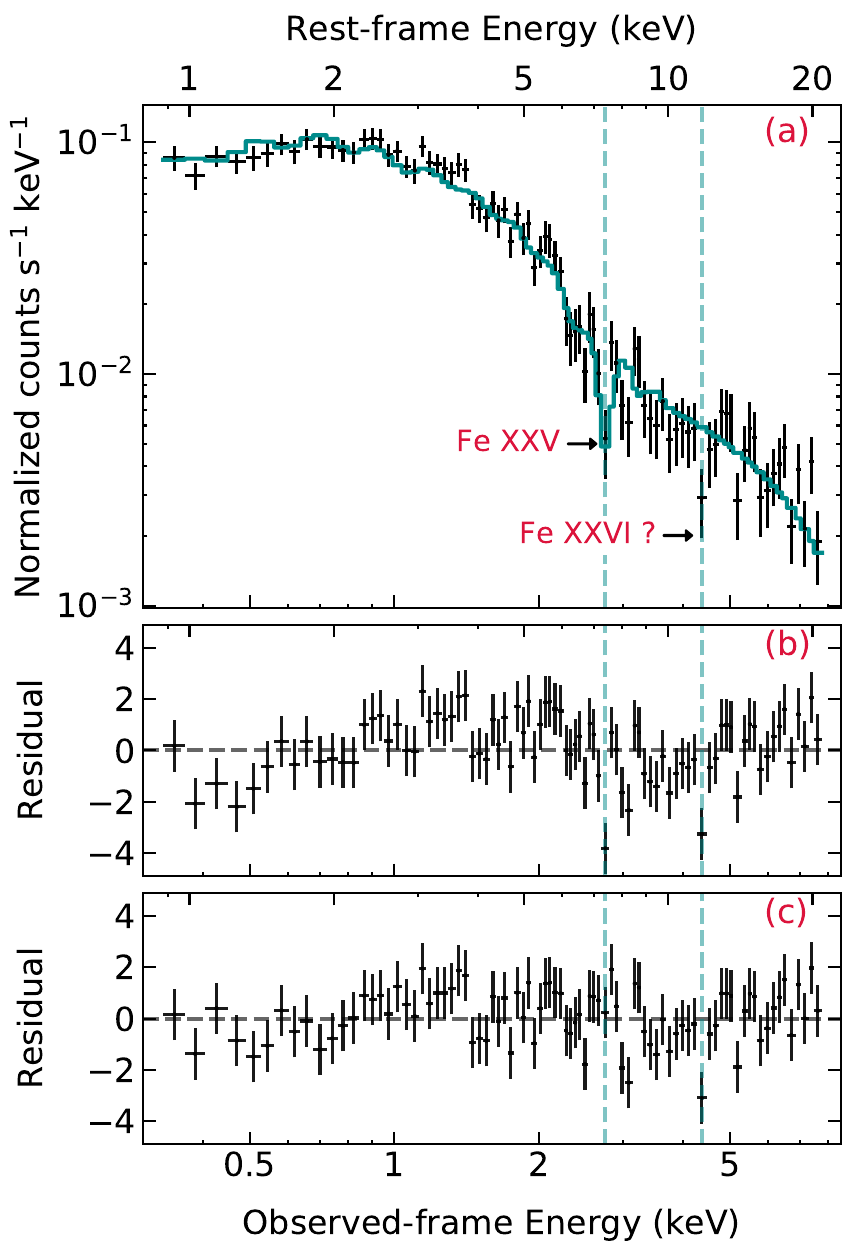}
	\caption{Panel (a): XMM 2002 data (black) and best-fit model (solid blue line) with Model pl\_wa. The dashed blue lines indicate the UFO signatures. Panel (b): XMM 2002 best-fit residuals for Model {\sf pl}. Panel (c): XMM 2002 best-fit residuals for Model {\sf pl\_wa}. The latter model self-consistently accounts for the absorption line at $E_{\rm rf}\simeq7.4\ {\rm keV}$. The data are grouped so to obtain at least 20 cts/bin, with minimum energy width set to one third of the CCD energy resolution. The best-fit parameters are summarized in Table \ref{tab:xmm_2002_abs_recap}. Due to background-dominated bins above ${\rm8.0\ keV}$, we restricted the fitting to the 0.3--8 keV observed-energy range (0.8--22 keV rest-frame energy range). }
	\label{fig:xmm_2002_cfr_del}
\end{figure}

The observed-frame best-fit residuals to Model {\sf pl} (Fig. \ref{fig:xmm_2002_cfr_del}, panel b and Table \ref{tab:xmm_2002_abs_recap}) show clear spectral complexities throughout the whole energy band. Those catching the eye are a deficit of counts at ${\rm\sim2.7\ keV}$ ($E_{\rm rf}\sim7.4\ {\rm keV}$) and at ${\rm\sim4.4\ keV}$ ($E_{\rm rf}\sim11.8\ {\rm keV}$), the latter being consistent to the majority of those found in the \textit{Chandra} spectra. Moreover, an excess of counts at ${\rm\sim2.1\ keV}$ (${\rm E_{rf}\sim5.7\ keV}$) and signatures of absorption in the soft band below ${\rm\sim0.6\ keV}$ ($E_{\rm rf}\sim1.6\ {\rm keV}$) are also present. 

Based on the results obtained from the \textit{Chandra} observations, we investigated the need for an extra cold absorber at the quasar's redshift, either uniformly or partially covering the source (Model {\sf pl\_a}={\tt phabs*zphabs*zpo}; and Model {\sf pl\_pca}={\tt phabs*zpcf*zpo}, respectively; best-fit parameters in Table \ref{tab:xmm_2002_abs_recap}). The data do require extra absorption at the quasar redshift and are slightly better (${\rm\Delta\chi^2\sim6}$ for one additional parameter) reproduced by a partial-covering medium. 
By adding a narrow Gaussian component to Model {\sf pl\_pca} for each of the narrow features hinted by the residuals in Fig. \ref{fig:xmm_2002_cfr_del} (panel b), the two in absorption turn out to be required by the data (${\rm\Delta\chi^2>10}$ for the addition of two parameters, each), while the emission line is only marginally detected (90\% confidence).
\renewcommand{\arraystretch}{1.15}
\begin{table*}[h]
	\caption{Summary of the best-fit parameters for each model tested for XMM 2002 data. }
    \centering
	\begin{tabular}{lcccccccccc}
	\hline\hline
		Model &${\rm \Gamma}$&$N{\rm_H}$&CF&${\rm log\xi}$&$z{\rm_{abs}}$&R&$E{\rm _{e.l.}}$&${\rm \Delta \chi^2}$&${\rm \Delta\nu}$ & ${\rm\chi^2_r\ (\nu)}$ \\ \hline 
		 {\sf pl} & ${\rm 1.66\pm0.04}$ & -- & -- & -- & -- & -- & -- & -- & -- & 1.56 (85) \\ 
		 {\sf pl\_a} & ${\rm 1.82\pm0.07}$ & ${\rm 0.32\pm0.12}$ & -- & -- & 1.695 & -- & -- & 22.6 & 1 & 1.31 (84)\\ 
		 {\sf pl\_pca} & ${\rm 1.91\pm0.11}$ & ${\rm2.1_{-1.2}^{+1.4}}$ & ${\rm 0.50_{-0.12}^{+0.10}}$ & -- & 1.695 & -- & -- &  28.9 & 2 & 1.25 (83)\\ 
		 {\sf pl\_wa} & ${\rm 1.74\pm0.04}$ & ${\rm 28.2_{-1.8}^{ +1.4}}$ & -- & ${\rm 2.5_{-0.3}^{+0.1}}$ & ${\rm\simeq1.445}$ & -- & -- & 34.1 & 3 & 1.20 (82) \\ 
		 {\sf pl\_pca\_pex\_el} & ${\rm 2.57\pm0.32}$ & ${\rm 2.3_{-0.7}^{ +0.7}}$ & ${\rm 0.74_{-0.11}^{+0.07}}$ & -- & 1.695 & ${\rm 0.58_{-0.27}^{+0.40}}$ & ${\rm5.7_{-0.2}^{+0.2}}$ & 54.2 & 5 & 0.98 (80) \\ 
		 {\sf pl19\_pca\_pex\_el} & ${\rm 1.90}$ & ${\rm 1.1_{-0.8}^{ +1.0}}$ & ${\rm 0.51_{-0.12}^{+0.44}}$ & -- & 1.695 & ${\rm <0.18}$ & ${\rm 5.7_{-0.2}^{+0.2}}$ & 41.1 & 4 & 1.13 (81)\\ 
		\hline
	\end{tabular}
	\label{tab:xmm_2002_abs_recap}
	\tablefoot{Col. 1: model name; Col. 2: photon index; Col. 3: column density in excess to the Galactic value (units of ${\rm 10^{22}\ cm^{-2}}$); Col. 4: covering fraction of the extra absorption; Col. 5: logarithm of the extra-absorption ionization parameter (${\rm erg\ s^{-1}\ cm}$); Col. 6: observed redshift of the extra absorption; Col. 7: reflection scaling factor; Col. 8: energy of the emission line (units of keV); Col. 9, 10: ${\rm \Delta \chi^2}$, ${\rm \Delta \nu}$ w.r.t Model {\sf pl} (${\rm \chi^2=132.6, \nu=85}$); Col. 11: ${\rm\chi^2_r}$. The energy width of the emission line is set to 0.01 keV.  All the errors are computed at 90\% confidence level for one parameter of interest. \textit{Model list}: Model {\sf pl} = {\tt phabs*zpo}; Model {\sf pl\_a} = {\tt phabs*zphabs*zpo}; Model {\sf pl\_pca} = {\tt phabs*zpcf*zpo};Model {\sf pl\_wa} = {\tt phabs*warmabs*zpo}; Model {\sf pl\_pca\_pex\_el} = {\tt phabs*zpcf*(zpo+pexrav+zgauss)}; Model {\sf pl19\_pca\_pex\_el} = {\tt phabs*zpcf*(zpo+pexrav+zgauss)} with ${\rm\Gamma=1.9}$. All the models include the Galactic absorption ($N{\rm_H=5.1\times10^{20}\ cm^{-2}}$). }
\end{table*}
\renewcommand{\arraystretch}{1.0}
Following \citet{protassov2002}, we evaluated the actual significance of the two narrow absorption features through Monte Carlo simulations, as follows. To obtain the null models, we started from Model {\sf pl\_pca} plus the three narrow Gaussian components; the absorption line whose actual significance was to be measured was then deleted before performing the simulations.
For each of the two null models, we simulated 1000 spectra through the {\tt fakeit} function in {\tt Xspec}. The two absorption lines turn out to be detected at $E{\rm_{rf}=7.4\ keV}$ and $E{\rm_{rf}=11.8\ keV}$ at confidence levels of 97.9\% and 87\%, respectively. 

Given their energies, they are likely blueshifted iron resonant lines, thus produced by a (highly) ionized and outflowing material. To test this scenario, we decided to use {\tt warmabs} \citep{kallman2001}, an analytic XSTAR model for self-consistent ionized absorption (Model {\sf pl\_wa}={\tt phabs*warmabs*zpo}), which takes into account also the production of absorption lines. Based on the results for measurements in other high-$z$ lensed quasars (Chartas et al., in prep.), we assumed Solar abundances and a gas turbulent velocity of ${\rm5000\ km\ s^{-1}}$, letting the column density ($N{\rm_H}$), the ionization parameter (${\rm log\xi}$) and the redshift of the absorber ($z{\rm_o}$) to vary. The {\tt warmabs} model assumes that the AGN emission modeled in {\tt Xspec} is the very same radiation that illuminates and ionizes the absorber. The initial conditions (abundances and density) of the medium are loaded through a population file, which depends on the power-law slope of the illuminating radiation. To find the best-fit $\Gamma$, we created different population files ($n{\rm =4\times10^4\ cm^{-3}}$, $v{\rm_{turb}=5000\ km\ s^{-1}}$) until the incident-radiation and the best-fit power-law photon indexes converged (within the errors). Model {\sf pl\_wa} yielded the best description of the XMM 2002 data (${\rm\chi^2_r=1.20}$) and best-fit parameters for the ionized absorber of: $N{\rm_H=2.8\pm0.2\times 10^{23}\ cm^{-2}}$, ${\rm log\left(\xi/erg\ s^{-1} cm\right)=2.5_{-0.3}^{+0.1}}$ and $z{\rm_o\simeq1.445}$ (see Table \ref{tab:xmm_2002_abs_recap} and Fig. \ref{fig:xmm_2002_cfr_del}, panel a and c).
The observed value $z{\rm_o}$ of the absorber redshift is related to the intrinsic redshift $z{\rm_a}$ of the medium (i.e. in the source rest frame) as $\left(1+z_{\rm o}\right)=\left(1+z_{\rm a}\right)\left(1+z_{\rm q}\right)$. Thus, the outflow velocity $v{\rm_{out}}$ can be determined from the relativistic Doppler effect formula: $1+z_{\rm a}=\sqrt{(1-\beta)/(1+\beta)}$, where $\beta=v_{\rm out}/c$. Given $z_{\rm q}=1.695$, the outflow velocity corresponding to $z_{\rm o}\simeq1.444$ is $v_{\rm out}=0.10\pm0.01c$. This ionized wind model naturally explains both the structure in the soft band and the absorption feature at $E{\rm_{rf}\simeq7.4\ keV}$ (Fig. \ref{fig:xmm_2002_cfr_del}), as opposed to the absorber of Model {\sf pl\_pca}, which, being cold, cannot originate such line. Given the wind ionization state and the outflow velocity, this line is consistent with being dominated by \ion{Fe}{xxv}, that has a rest-frame energy at rest of 6.7 keV. 
However, this same model fails to account for the second absorption line at $E{\rm_{rf}\simeq11.8\ keV}$ (${\rm 4.4\ keV}$, see Fig. \ref{fig:xmm_2002_cfr_del}, panel a and c). 
Moreover, the narrow emission line at $E{\rm_{rf}=5.7\pm0.2\ keV}$ is only marginally detected (90\% confidence) when adding a Gaussian component to Model {\sf pl\_wa}. Its energy is consistent with that of the line we find in \textit{Chandra} ObsID 431 A and with that of the microlensed Fe K$\alpha$ found by \citet{dai2003}. Its width (${\rm\sigma<0.01\ keV}$), however, is not consistent with the broad one found by \citet{dai2003} (${\rm\sigma=0.87_{-0.15}^{+0.30}\ keV}$). 
The results of \citet{fedorova2008}, who analyzed the XMM 2002 observation, tentatively detecting an emission line at $E{\rm_{rf}=6.0_{-1.0}^{+0.7}\ keV}$ with an energy-width upper limit of ${\rm\sigma<0.9\ keV}$, are consistent with ours.  

Finally, we stress that the significance obtained for the ${\rm7.4\ keV}$ absorption line through the Monte Carlo simulations does not correspond to that of the UFO. In fact, assessing the F-test significance of the {\tt warmabs} component leads to a detection of the outflow that is well above the 99.99\% confidence level. This is linked to the fact that the UFO does not only explain the absorption line by itself but also acts on the shape of the soft-band continuum, as the two lower panels of Fig. \ref{fig:xmm_2002_cfr_del} show. 

For the sake of completeness, we also tested a reflection scenario. Instead of self-consistent reflection models which bind the Fe K$\alpha$ line to the 6.4--6.7 keV range, we built a phenomenological model so to let the emission line be placed at lower energies (Model {\sf pl\_pca\_pex\_el}= {\tt phabs*zpcf*(zpo+pexrav+zgauss)}). The reflected-power-law photon index was set to that of the intrinsic emission, the abundances equal to Solar, the inclination angle to ${\rm60\degr}$, the cutoff energy to 300 keV and the reflection fraction to be negative, so to only model the reflected emission through the {\tt pexrav} component. The best-fit reflection fraction and photon index (Table \ref{tab:xmm_2002_abs_recap}) are ${\rm R=0.58_{-0.27}^{+0.40}}$ and ${\rm\Gamma=2.57\pm0.32}$ (${\rm\chi^2_r=0.98}$), which corresponds to a considerably steeper power law, at the limits of the expected values for an AGN \citep[e.g.][]{perola2002,piconcelli2005}. In this case, the emission line is detected at $E_{\rm rf}=5.7\pm0.2\ {\rm keV}$ above 99\% confidence as a narrow line, with an equivalent width of ${\rm EW=154_{-103}^{+99}\ eV}$. Following \citet{makishima1986}, this agrees with what is expected given the column density. However, this model is almost in tension with \citet{leahy2001} for what concerns the reflection fraction R. The steep power-law may be caused by the known photon index - column density degeneracy. Therefore, we tried setting the photon index (Model {\sf pl19\_pca\_pex\_el}) to the standard value ${\rm \Gamma=1.90}$ for high-$z$ quasars \citep[e.g.][]{vignali2005,just2007}, which is also consistent to the average ${\rm\Gamma}$ of the \textit{Chandra} HSS and that of the absorption models for this spectrum (see Table \ref{tab:xmm_2002_abs_recap}). When doing so, the reflection fraction becomes consistent with zero (90\% confidence upper limit: ${\rm R<0.18}$) and the quality of the fit decreases (${\chi^2_r=1.13}$). Therefore, the reflection component turns out not to be physically, in the first case, nor statistically, in the second case, required by the 2002 data, confirming also earlier results by \citet{fedorova2008}. 

\subsection{XMM 2018}
\begin{figure}[!h]
	\centering
	\includegraphics[width=0.97\linewidth]{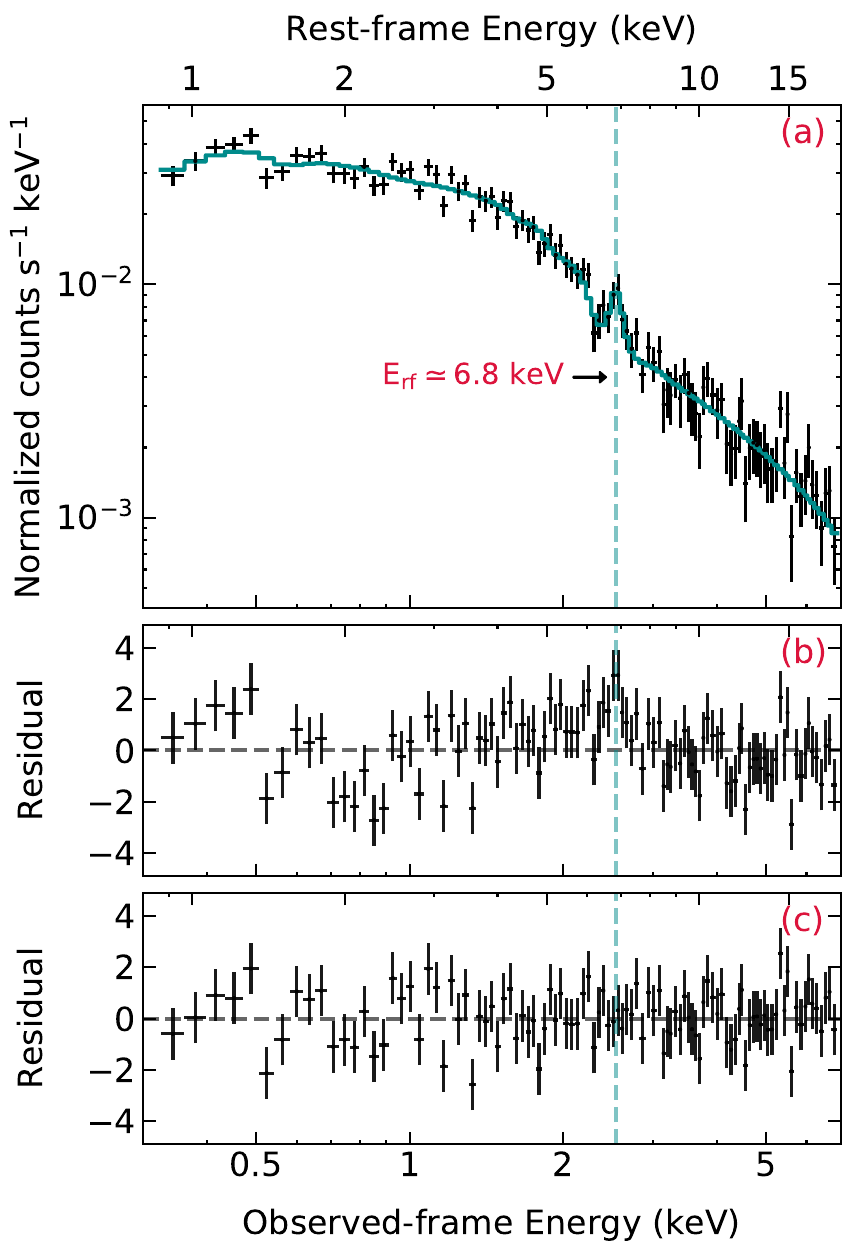}
	\caption{Panel (a): XMM 2018 data (black) and best-fit model (solid blue line) with Model pl\_pca\_el. The dashed blue line indicate the emission line at $E_{\rm rf}=6.84\pm0.11\ {\rm keV}$. Panel (b): XMM 2018 best-fit residuals for Model {\sf pl}. Panel (c): XMM 2018 best-fit residuals for Model {\sf pl\_pca\_el}. The data are grouped so to obtain at least 20 cts/bin, with minimum energy width set to one third of the CCD energy resolution. The best-fit parameters are summarized in Table \ref{tab:xmm_2018_recap}. Due to background-dominated bins above ${\rm7.0\ keV}$, we restricted the fitting to the 0.3--7.0\ keV observed-energy range (1--19 keV rest-frame energy range). }
	\label{fig:xmm_2018_cfr_del}
\end{figure}

As for the other spectra, we began our analysis of the XMM 2018 EPIC-pn spectrum by inspecting the best-fit residuals to Model {\sf pl} (Fig. \ref{fig:xmm_2018_cfr_del}, panel b, Table \ref{tab:xmm_2018_recap}). Due to background-dominated bins above 7.0 keV, we restricted the fitting to the 0.3--7.0 keV observed-energy range (1--19 keV rest-frame energy range). The residuals (Fig. \ref{fig:xmm_2018_cfr_del}, panel b) show complexities in the soft-energy band that are likely due to an absorber and indicate a prominent emission line just below ${\rm 3\ keV}$ in the observed frame. 
At higher energies, however, the distribution is quite flat, although noisy, suggesting the absence of a significant reflection component. No hints of absorption lines in the hard-energy band are seen either. 

Using the same logical steps as for XMM 2002, we started by adding more complex absorption models to fit the low energy continuum. All the best-fit parameters of the tested models are summarized in Table \ref{tab:xmm_2018_recap}. Given the shape of the residuals, the absorber during this observation could either be cold and partially covering the emitting source, or ionized (Model {\sf pl\_pca} and Model {\sf pl\_wa}\footnote{Chemical abundances and the gas turbulent velocity were set as done for XMM 2002 (see Sect. \ref{sec:xmm_2002}). The best-fit $\Gamma$ was found using the same method as for XMM 2002. }, respectively). For completeness, we also investigated the case of a cold medium blocking all the intrinsic emission (Model {\sf pl\_a}) which, as expected, turned out not to be required by the data. In all three cases, we set the absorber's redshift to the systemic of the quasar, based on the results obtained with the \textit{Chandra} data (Sect. \ref{sec:cha_continuum}) and because no absorption lines above ${\rm7\ keV}$ rest-frame were found (i.e. there are no hints of outflowing material). 
When limiting the analysis to one feasible absorption component, the XMM 2018 spectrum is best reproduced by a power-law emission modified by a partial covering medium (Model {\sf pl\_pca}, see Table \ref{tab:xmm_2018_recap}). 
The prominent emission line indicated by the residuals (Fig. \ref{fig:xmm_2018_cfr_del}, panel b) is found to be narrow (Model {pl\_pca\_el}, Table \ref{tab:xmm_2018_recap} and Fig. \ref{fig:xmm_2018_cfr_del}, panel a and c) with a 90\% energy-width upper limit ${\rm\sigma<0.53\ keV}$. Thus we find for XMM 2018 a rather thick absorber that blocks part of the intrinsic emission ($N{\rm_H\simeq1.0\times10^{23}\ cm^{-2}}$, ${\rm CF\simeq0.53}$) and a significant emission line with rest-frame energy and equivalent width of $E_{\rm  rf}=6.84\pm0.11\ {\rm keV}$ and ${\rm EW=267\pm111\ eV}$. The energy is inconsistent with both that of the marginal detection in the 2002 data (see Sect. \ref{sec:xmm_2002}) and that of the skewed emission line found by \citet{reynolds2014}. 
To see whether the production of this line could be ascribed to the absorber, we evaluated the upper limit of the medium ionization state through the {\tt warmabs} model. To mimic the partial absorption of the intrinsic emission, we included two power-law plus emission-line components, the first seen directly and the other as scattered by the absorber (Model {\sf pl\_pcwa}={\tt phabs*[(zpo+zga)+warmabs*(zpo+zga)]}), both modified by Galactic absorption. The slopes and the line parameters of the two terms were linked to each other, since the primary emission is the same for both components. We inferred a 90\% upper limit to the ionization parameter of ${\rm log\left(\xi/erg\ s^{-1} cm\right)=2.2}$ and a covering fraction ${\rm CF=0.58\pm0.33}$, which is consistent with that obtained with Model {\sf pl\_pca} (Table \ref{tab:xmm_2018_recap}). The inferred ionized state is not consistent to that of the UFO detected in XMM 2002, and nor are the column densities. Thus, we find unlikely for the two absorbers to be the same gas that changed in covering fraction. 

Following \citet{makishima1986}, the absorber ionization state (${\rm log\left(\xi/erg\ s^{-1} cm\right)\leq2.2}$) translates into a medium that is dominated by iron from \ion{Fe}{i} to \ion{Fe}{xx}, while, from its energy, the line we detect is consistent with being dominated by \ion{Fe}{xxv--xxvi}. Thus, this feature cannot be produced by the absorber itself since it would require a much more ionized gas (${\rm log \left(\xi/erg\ s^{-1} cm\right)\geq3.0}$). Unfortunately, we are not able to verify whether an additional highly ionized medium could be required by the data because the SNR of the present data does not allow us to constrain such a complex model. 
Another possible explanation for the ${\rm 6.8\ keV}$ line could be the microlensing differential magnification of a relativistic blurred Fe K$\alpha$ produced by the Compton reflection in the accretion disk, as proposed in other sources by \citet{chartas2016_gravlens,chartas2017}. Based on this argument, we tested whether a reflection scenario would give a better representation of the XMM 2018 data, using the same phenomenological model discussed in section \ref{sec:xmm_2002} (Model {\sf pl\_pca\_pex\_el}). This model returns a best fit that on a statistical basis is as good as that of Model {\sf pl\_pca} but its reflection fraction R is consistent with zero at the 90\% confidence level (${\rm R\leq0.16}$, see Table \ref{tab:xmm_2018_recap}).
On the one hand, this result confirms that the Compton reflection is not a dominant component in the Einstein Cross emission but, on the other hand, it does not completely rule out the interpretation of the ${\rm6.8\ keV}$ emission line as a microlensed Fe K$\alpha$ line. Standard reflection models, as {\tt pexrav}, assume the reflection continuum as produced by the whole disk. In the case of a microlensing event magnifying the inner regions of the approaching side of the accretion disk, \citet{popovic2006} demonstrate that only the emission associated to the blueshifted part of the Fe K$\alpha$ line is enhanced, while the reflection continuum is not, unless the microlensing event is monitored for its whole duration. Since the source crossing time in the Einstein Cross is estimated to be of a few months \citep{mosquera_kochanek_2011}, we cannot rule out the possibility of this emission line to be a microlensed Fe K$\alpha$. 

\renewcommand{\arraystretch}{1.15}
\begin{table*}[!h]
	\caption{Summary of the best-fit parameters for each tested model for XMM 2018 EPIC-pn spectrum. The fitting was carried out over the 0.3--7.0 keV observed-energy range because the data is background dominated above 7 keV. }
	\centering
	\begin{tabular}{lcccccccccc}
		\hline\hline
		Model&$\Gamma$&$N{\rm_H}$&CF&${\rm log\xi}$&$z_{\rm abs}$&R&$E{\rm_{e.l.}}$&${\rm \Delta \chi^2}$&${\rm \Delta\nu}$& ${\rm\chi^2_r\ (\nu)}$ \\  
		\hline
		{\sf pl} & ${\rm 1.56\pm0.03}$ & -- & -- & -- & -- & -- & -- & -- & -- & 1.64 (102) \\ 
		{\sf pl\_a}& ${\rm 1.56\pm0.03}$ & ${\rm <0.04}$ & -- & -- & 1.695 & -- & -- & -0.4 & 1 & 1.72 (101)\\ 
		{\sf pl\_pca} & ${\rm 1.99\pm0.10}$ & ${\rm11.1_{-2.2}^{+2.5}}$ & ${\rm 0.56\pm0.08}$ & -- & 1.695 & -- & -- & 47.3 & 2 & 1.20 (100)\\ 
		{\sf pl\_pca\_el}& ${\rm 1.96\pm0.10}$ & ${\rm10.3_{-2.2}^{+2.5}}$ & ${\rm 0.53\pm0.09}$ & --  & 1.695 & -- &${\rm 6.84\pm0.11}$ & 62.4 & 4 & 1.07 (98)\\ 
		{\sf pl\_wa} &${\rm1.62\pm0.03}$ & ${\rm<0.19}$ & -- & ${\rm 2.4_{-0.6}^{+0.4}}$ & 1.695 & -- & -- & 2.1 & 2 & 1.65 (100) \\ 
		{\sf pl\_pca\_pex\_el}& ${\rm 2.03_{-0.15}^{+0.20}}$ & ${\rm9.4_{-2.3}^{+3.0}}$ & ${\rm 0.56\pm0.09}$ & -- & 1.695 &  ${\rm <0.16}$&${\rm 6.84\pm0.10}$ & 63.7 & 5 & 1.07 (97) \\ 
		\hline
	\end{tabular}
	\label{tab:xmm_2018_recap}
	\tablefoot{Col. 1: model name; Col. 2: photon index; Col. 3: column density in excess to the Galactic value (units of ${\rm 10^{22}\ cm^{-2}}$); Col. 4: covering fraction of the extra absorption; Col. 5: logarithm of the extra-absorption ionization parameter (${\rm erg\ s^{-1}\ cm}$); Col. 6: observed redshift of the extra absorption; Col. 7: reflection scaling factor; Col. 8: energy of the narrow emission line (units of keV); Col. 9, 10: ${\rm \Delta \chi^2}$, ${\rm \Delta \nu}$ w.r.t Model {\sf pl} (${\rm \chi^2=167.4, \nu=102}$); Col. 11: ${\rm\chi^2_r}$. The energy width of the emission line is set to 0.01 keV. All the errors are computed at 90\% confidence level for one parameter of interest. \textit{Model list}: Model {\sf pl} = {\tt phabs*zpo}; Model {\sf pl\_a} = {\tt phabs*zphabs*zpo}; Model {\sf pl\_pca} = {\tt phabs*zpcf*zpo}; Model {\sf pl\_wa} = {\tt phabs*warmabs*zpo} with $z\equiv z_{\rm q}$; Model {\sf pl\_pca\_el} = {\tt phabs*zpcf*(zpo+zgauss)}; Model {\sf pl\_pca\_pex\_el} = {\tt phabs*zpcf*(zpo+pexrav+zgauss)} with ${\rm\Gamma=1.9}$. All the models include the Galactic absorption (N${\rm_H=5.1\times10^{20}\ cm^{-2}}$). }
\end{table*}
\renewcommand{\arraystretch}{1} 
\subsection{Summary of the XMM-Newton results}
\begin{itemize}
    \item The XMM 2002 spectra are best physically and statistically reproduced by a complex absorber with $N{\rm_H\simeq2\times 10^{23} cm^{-2}}$, ${\rm log\left(\xi/erg\ s^{-1} cm\right)\simeq3.0}$, and outflow velocity of $v{\rm_{out}\sim0.1c}$, typical of UFOs. This also explains the prominent absorption iron resonant line measured at $E_{\rm rf}\simeq7.4\ {\rm keV}$ and interpreted as \ion{Fe}{xxv}. 
    \item The 2018 data do not show any similar blueshifted absorption features and are best fitted by a partial-covering mildly-ionized material, with $N{\rm_H\simeq1.0\times10^{23}\ cm^{-2}}$, ${\rm CF\simeq0.53}$ and with a 90\% ionization-state upper limit of ${\rm log \left(\xi/erg\ s^{-1} cm\right)<2.2}$. We detect a significant narrow emission line at $E_{\rm rf}=6.84\pm0.11\ {\rm keV}$, with an equivalent width of ${\rm EW=267\pm111\ eV}$. This line is in tension with being produced by the absorber itself because, given its energy, a much more ionized medium would be required. 
    \item Constraining the reflection component is challenging for both the XMM spectra, also due to the limited energy range provided at high energies. We find that for the 2002 data, a reflection component is statistically significant only when a very steep power-law (${\rm \Gamma\simeq2.65}$) is assumed, while it is negligible when a typical AGN slope (${\rm \Gamma\simeq1.9}$) is adopted. Regarding the 2018 data, despite the presence of a prominent emission line at $E_{\rm rf}=6.84\pm0.11\ {\rm keV}$, a reflection component is found not to be statistically required. 
\end{itemize}

\section{Discussion and results}
\label{sec:disc_res}
We have presented the results obtained from the analysis of all the available X-ray data of the Einstein Cross (Q 2237+030), a quasar at $z=1.695$ that is gravitationally lensed in four images by a foreground spiral galaxy. We analyzed 40 archival observations, 37 taken by \textit{Chandra} and three by XMM-\textit{Newton}, covering a period of 18 years, for a total of ${\rm\sim0.9\ Ms}$. 

From the \textit{Chandra} data, we probed the source spectral variability, using the photon-index variations through the epochs as proxy. These are qualitatively consistent among the four images (i.e. intrinsic), which supports the assumption made by \citet{chen2012}, who linked the photon index among the images when fitting spectra extracted from the same observation. 
To assess the origin of such variability, we limited the analysis to the HSS, i.e. the fourteen spectra extracted from 11 observations that show the highest number of counts (above 500 cts in the 0.4--7 keV observed-frame energy range), which allowed us to better constrain the model parameters. We find that an additional cold absorber is highly required (above 99\% confidence) in four of the HSS spectra, corresponding to three different epochs. Moreover, the column density is consistent with being variable at more than 99\% confidence between the epochs. Thus, the spectral variability is likely ascribed to a variable absorber placed at the quasar's redshift, but we cannot exclude that part of it could be produced by the variation of the source intrinsic power-law emission as well. 

The XMM-\textit{Newton} data are fundamental in investigating the need for extra absorption and the nature of the medium, given the much higher counting-statistics they provide. We find that the XMM 2002 data are consistent with a UFO scenario with $N{\rm_H=2.0_{-0.5}^{+0.6}\times 10^{23}\ cm^{-2}}$, ${\rm log\left(\xi/erg\ s^{-1} cm\right)=3.0\pm0.1}$ and $v_{\rm out}=0.1\pm0.01c$, that explains the prominent absorption line at $E_{\rm rf}= 7.4\pm0.1\ {\rm keV}$. However, the same UFO cannot explain the second absorption line detected at ${\rm\sim11.8\ keV}$, unless we assume it to be the hints of a blueshifted \ion{Fe}{xxvi} Ly${\rm\alpha}$ ($E_{\rm rf}=6.97\ {\rm keV}$ at rest) produced by an even faster component outflowing at $\sim0.5 c$. This would not be the first UFO showing more than one outflow component and at such extremely high velocities \citep[see, for instance, the case of APM 08279+5255,][]{chartas2009}. 

The rest-frame absorption-corrected 2--10 keV luminosity of Q2237 during the 2002 observation is $L{\rm_{2-10}\simeq 6.6\times10^{45}\ erg\ s^{-1}}$. Given a magnification factor of ${\rm\mu\approx16}$ \citep{schmidt1998}, the intrinsic absorption-corrected luminosity is $L{\rm_{2-10}^{int}\simeq 4.1\times10^{44}\ erg\ s^{-1}}$. 
From the UV luminosity $\log\left(\lambda L_{\lambda}\right)_{\rm 1450\AA{}}\simeq45.53$ reported in \citet{assef2011} and assuming a conversion factor of ${\rm\simeq4}$ \citep{richards_2006}, we find a bolometric luminosity of $L{\rm_{bol}\simeq 1.4\times10^{46}\ erg\ s^{-1}}$. 
Based on \citet{lusso2012} and the recent work by \citet{duras_2020}, the predicted 2--10 keV intrinsic luminosity would be $L{\rm_{2-10}^{int}\simeq 4\times10^{44}\ erg\ s^{-1}}$, which is in good agreement with the one we measure. \citet{assef2011} estimate the black hole mass $M{\rm_{BH}}$ from the H${\rm\beta}$ broadening to be $\log\left(M_{\rm BH}/M_{\sun}\right)=9.08\pm0.39$, which leads to an Eddington luminosity of $L_{\rm Edd}\simeq 1.5\times10^{47}\ {\rm erg\ s^{-1}}$ (${\rm\lambda_{Edd}\approx0.1}$). 

Assuming the high significance of the outflow observed in XMM 2002 at $v_{\rm out}\simeq0.1c$, we can derive the physical properties of the wind, by adopting standard 'prescriptions' \citep[e.g.][]{tombesi2012,crenshaw_2012} and including the uncertainties on the best-fit parameters, so to place this detection in a broader context and compare it with the measurements in other QSOs at $z\geq1.5$ and in the local Universe. 
Following \citet{crenshaw_2012}, the mass-outflow rate can be obtained using the formula $\dot{M}_{\rm out}=4\pi r\mu m_{\rm p} N_{\rm H} v_{\rm out}C_{\rm g}$, where $m_{\rm p}$ is the proton mass, $\mu$ is the mean atomic mass per proton (1.4 for solar abundances), $r$ is the distance from the BH and $C_{\rm g}$ is the global covering factor of the wind. We assume $C_{\rm g}\approx0.5$, based on the statistical study carried out by \citet{tombesi2010}, and recently confirmed by \citet{igo_2020}, over a sample of local Seyfert galaxies. Moreover, we conservatively assume the outflow to be detected at the minimum distance from the BH, where the observed velocity $v{\rm_{out}}$ equals the escape velocity from the BH potential well, thus $r_{\rm min}=2{\rm G}M_{\rm BH}/v_{\rm out}^{\ 2}$. We obtain $r{\rm_{min}\approx3.6\times 10^{16}\ cm}$, that corresponds to ${\rm\simeq 100}$ gravitational radii ($r_{\rm g}={\rm G}M_{\rm BH}/c^2$); considering the uncertainties on $M_{\rm BH}$, we find for $r_{\rm min}$ the range $r{\rm_{min}\approx(0.9-7)\times 10^{16}\ cm}$. 
Using $r=r_{\rm min}$ as radial location of the outflow, we are estimating the lower limit to the following quantities. The mass outflow rate, given all the assumptions above, turns out to be $\dot{M}_{\rm out}\approx5\ M_{\sun}\ {\rm yr^{-1}}$. Taking into account the 1${\rm\sigma}$ uncertainties of the parameters, we find quite a wide range for the mass-outflow rate: ${\rm \dot{M}_{out}\sim(0.6-10.3)\ M_{\sun}\ yr^{-1}}$. As a result, all the following quantities derived using $\dot{M}_{\rm out}$ will have likewise wide uncertainties. 
The outflow mechanical output ($\dot{E}_{\rm kin}=\frac{1}{2}\dot{M}_{\rm out}v_{\rm out}^2$) is $\dot{E}_{\rm kin}=1.5\times10^{45}\ {\rm erg\ s^{-1}}$, which corresponds to an outflow efficiency of $\dot{E}_{\rm kin}/L_{\rm bol}\approx0.1$. We obtain an outflow momentum rate of $\dot{p}_{\rm out}=\dot{M}_{\rm out}v_{\rm out}\approx 9.9\times10^{35} {\rm cm\ g\ s^{-2}}$, that is approximately twice the radiation pressure $\dot{p}_{\rm rad}= L_{\rm bol}/c$. Therefore, this UFO is consistent with generating efficient wind-driven AGN feedback that might indeed act on the evolution of the quasar host galaxy, given the $\dot{E}_{\rm kin}/L_{\rm bol}>0.5$\%--5\% threshold predicted by the models \citep[e.g.][]{dimatteo2005,hopkins2010}. 
Moreover, being the outflow momentum rate higher than $L_{\rm bol}/c$, magnetic forces might be playing a non-secondary role in accelerating this UFO. 
The derived wind parameters (column density, ionization state and outflow velocity) are consistent with those of UFOs in local AGN \citep{tombesi2010} but the kinematic properties, albeit the wide uncertainties, seem to be higher than the average values for local objects \citep{tombesi2012}. They are instead consistent with those of high-$z$ AGN, for instance of PID352 \citep{vignali2015}, a bright, unlensed source at $z\approx1.6$ that shows a similar $L_{\rm bol}$ (${\rm\sim10^{46}\ erg\ s^{-1}}$). Furthermore, the properties of this UFO agree with the $v_{\rm out}-L_{\rm bol}$ and $L_{\rm bol}-\dot{E}_{\rm kin}$ relations in \citet{fiore2017}, computed for a compilation of local and (few) high-redshift X-ray winds. 

Interestingly, the XMM 2018 spectra do not show any absorption line in the hard band and seem to be best reproduced by a partial-covering mild-ionized absorber, with ionization parameter of ${\rm \log\left(\xi/erg\ s^{-1} cm\right)\leq 2.2}$ (90\% confidence limit). The intrinsic, absorption-corrected 2--10 keV luminosity for the 2018 observation is $L{\rm_{2-10}^{int}\simeq2.0\times10^{44}\ erg\ s^{-1}}$, approximately 49\% of that found for the 2002 data. From the upper limit to the ionization state, we evaluated the lower limit to the absorber maximum distance from the central BH (being $\xi=L_{\rm X}/N_{\rm H}r_{\rm max}$): $r_{\rm max}\geq4.7\ {\rm pc}$. Thus, it seems to be placed at a distance consistent with the typical range of the BLR or the molecular torus \citep[e.g.][]{jaffe2004,burtscher2013}. Since accretion-disk winds are thought to have a global covering fraction less than unity, we propose a scenario where the wind has changed its direction w.r.t. the los and the disk between the two XMM observations, and part of the clouds contained within the molecular torus or the BLR intercept the los during the second pointing. Given the short time elapsed, we think it is unlikely that the outflow has been totally suppressed between the two observations. Given that the \textit{Chandra} observations were taken in between XMM 2002 and XMM 2018, we find that our statement is supported by the UFO signatures in the \textit{Chandra} data. Moreover, the lowest timescale of the absorber variability obtained from \textit{Chandra} HSS (${\rm\simeq0.3\ yrs}$ rest frame) is linked to its distance from the central BH as $d\approx c\Delta t\approx0.09\ {\rm pc}$, consistent with the innermost regions of the torus and of the BLR, thus consistent with the proposed scenario \citep[e.g.][]{perola2002,risaliti2009,burtscher2013}. However, the lower statistics of the \textit{Chandra} spectra did not allow us to investigate the presence of more complex absorbers than a neutral medium, i.e. to model the absorption lines we detect with a wind model producing them. 

In the \textit{Chandra} HSS we find emission lines that span from ${\rm\sim2.2}$ to ${\rm\sim6.5\ keV}$ rest frame, which, following \citet{dai2003} and \citet{chartas2016_gravlens,chartas2017}, might be interpreted as microlensed and relativistic Fe K$\alpha$ lines. Moreover, the energy range they cover is consistent with the energies of the redshifted Fe K$\alpha$ emission lines found by \citet{chartas2016_gravlens} in RX J1131-1231. Interestingly, the only highly significant emission line is consistent with a regular Fe K$\alpha$ line ($E{\rm_{rf}=6.47_{-0.12}^{+0.11}\ keV}$). Despite this fact, also this line seems to be microlensed since we do not detect it in the stacked spectrum of images B+C+D from the same observation. From a preliminary search for microlensing events in the \textit{Chandra} multi-epoch light-curve ratios, there is no clear link between the observations in which we detect these emission lines and microlensing effects. In XMM 2018 we find a significant emission line at ${\rm\simeq6.8\ keV}$ that is in tension with being produced by the absorber, due to its ionization state. Even though a reflection component is not required by the data, such a line could be produced by a microlensing caustic that is crossing the inner regions of the accretion-disk's approaching side. Such a microlensing event will lead to the magnification of blueshifted Fe K$\alpha$ emission from a narrow inner region of the disk without magnifying the distant reflected continuum \citep{popovic2006,Krawczynski_2017,Krawczynski_2019}. The present data, however, do not allow us to verify either the presence of another highly ionized absorption component, since the data SNR is too low to constrain such a complex model, or the case of a microlensing event magnifying the inner regions of the accretion disk, since longer and less sparse \textit{Chandra} monitoring would be required. 
 
In the \textit{Chandra} spectra, we detect, for the first time in this source, several blueshifted iron resonant absorption lines, with overall significance slightly below 3$\sigma$. 
Interestingly, all the most significant lines (${\rm>99\%}$ confidence level) of the HSS (Table \ref{tab:cha_high_lines_abs}) are grouped around the value of ${\rm11\ keV}$ and have energies consistent to the least significant (87\% confidence level) feature of XMM 2002. If confirmed, they would imply a second and more extreme wind component with $ v_{\rm out}\approx0.3$--$0.5c$. Thus, the Einstein Cross could have experienced a multi-velocity UFO event, as found for other quasars, either nearby (e.g. PDS 456, \citealt{Boissay-Malaquin2019}; IRAS 00521--7054, \citealt{walton_2019}) or more distant (e.g. APM 08279+5255, \citealt{chartas2009}). 

Given the large number of X-ray observations of Q2237, we could roughly evaluate, for the first time ever, the wind duty cycle for this single source. However, almost two thirds of the dataset did not provide a SNR high enough to constrain the presence/absence of the narrow features.
In total, we detect UFO signatures at more than 90\% confidence in six observations (5 from \textit{Chandra}, 1 from XMM-\textit{Newton}) out of the thirteen analyzed to this purpose (11 from \textit{Chandra}, 2 from XMM-\textit{Newton}). Thus, we find the wind duty cycle to be $DC_{\rm w}\approx0.46$ at 90\% confidence. If we only consider those observations showing absorption lines at more than 95\% confidence (3 from \textit{Chandra}, 1 from XMM-\textit{Newton}), the duty cycle turns out to be $DC_{\rm w}\approx0.31$. Nonetheless, we cannot exclude the presence of UFO signatures that are too weak to be detected due to the too low SNR of the HSS spectra showing the least number of counts. For this reason, our estimates of $DC_{\rm w}$ represent the wind-duty-cycle lower limit over the thirteen observations that provide data with high-enough statistics. 
Although strictly related to the signal-to-noise of the spectra, our estimation of this parameter represents the best we are able to achieve with present-day data.

With this work we are adding a new piece to the puzzle of high-redshift AGN ($z>1.5$) which do show complex spectra and sometimes show variable UFOs. So far, only around ten objects at $z\geq1.5$ were analyzed to this purpose (of which only two are non-lensed quasars: HS 1700+6416, \citealt{lanzuisi2012}; PID 352, \citealt{vignali2015}), leading to a detection fraction of ${\rm\sim70\%}$ (Chartas et al. in prep). 
Expanding this high-redshift sample will be key in the future to better assess the occurrence and the properties of these events at the peak of the QSO activity. To this aim, the launch of new-generation X-ray observatories (i.e. \textit{Xrism} and \textit{Athena}) will provide us the means to obtain spectra with much higher signal-to-noise ratios and spectral resolution w.r.t those we can reach nowadays, essential ingredient in unveiling UFOs. This would allow us to better understand if and how disk-driven winds do trigger efficient AGN feedback at a time where the scaling relations between SMBHs and host galaxies were put in place. 

\begin{acknowledgements}
    The authors thank the referee for the careful reading of the manuscript and the helpful comments. The authors also thank Gabriele Ponti for useful discussions. 
    This work is based on observations obtained with XMM-Newton, an ESA science mission with instruments and contributions directly funded by ESA Member States and the USA (NASA). This research has made use of data obtained from the \textit{Chandra} Data Archive, and software provided by the \textit{Chandra} X-ray Center (CXC) in the application package CIAO. 
    CV and MD acknowledge financial support from the Italian Space Agency (ASI) under the contracts ASI-INAF I/037/12/0 and ASI-INAF n.2017-14-H.0. 
    BDM acknowledges support from the European Union’s Horizon 2020 research and innovation programme under the Marie Skłodowska- Curie grant agreement No. 798726. 
    MG is supported by the ''Programa de Atracci\'on de Talento'' of the Comunidad de Madrid, grant 2018-T1/TIC-11733. 
\end{acknowledgements}

\bibliographystyle{aa}
\bibliography{ebert_q2237}

\begin{appendix}
\section{Chandra stacked spectra}
\label{app:cha}
After analyzing the spectral properties of each \textit{Chandra} observation, we inspected the stacked spectra to see the persistence of the features we find in the \textit{Chandra} data and how these compare to those in the XMM-\textit{Newton} data. First, we produced the individual-image stacked spectra combining all the epochs (Appendix \ref{app:single_ima_stacked}), then we combined all the images from all the epochs to obtain the final stacked spectrum (Appendix \ref{app:all_combined}). 
All the spectra were combined through the CIAO tool {\tt combine\_spectra}, then grouped so to obtain at least 20 cts/bin and analyzed applying the ${\rm\chi^2}$ statistics.  All the best-fit values to each tested model are summarized in Table \ref{tab:cha_stack}.

\subsection{Stacked spectra of the individual images}
\label{app:single_ima_stacked}
The single-image stacked spectra (total source time of ${\rm 749\ ks}$ per each image) of images B, C and D show a similar number of counts in the 0.4--7 keV observed-energy band (5500\,cts, 3975\.cts and 4766\,cts, respectively), comparable to that of XMM 2002 and 2018, while for image A we obtain a much better statistics (17700\,cts). 

\begin{figure}[h]
			\centering
			\includegraphics[width=1.0\linewidth]{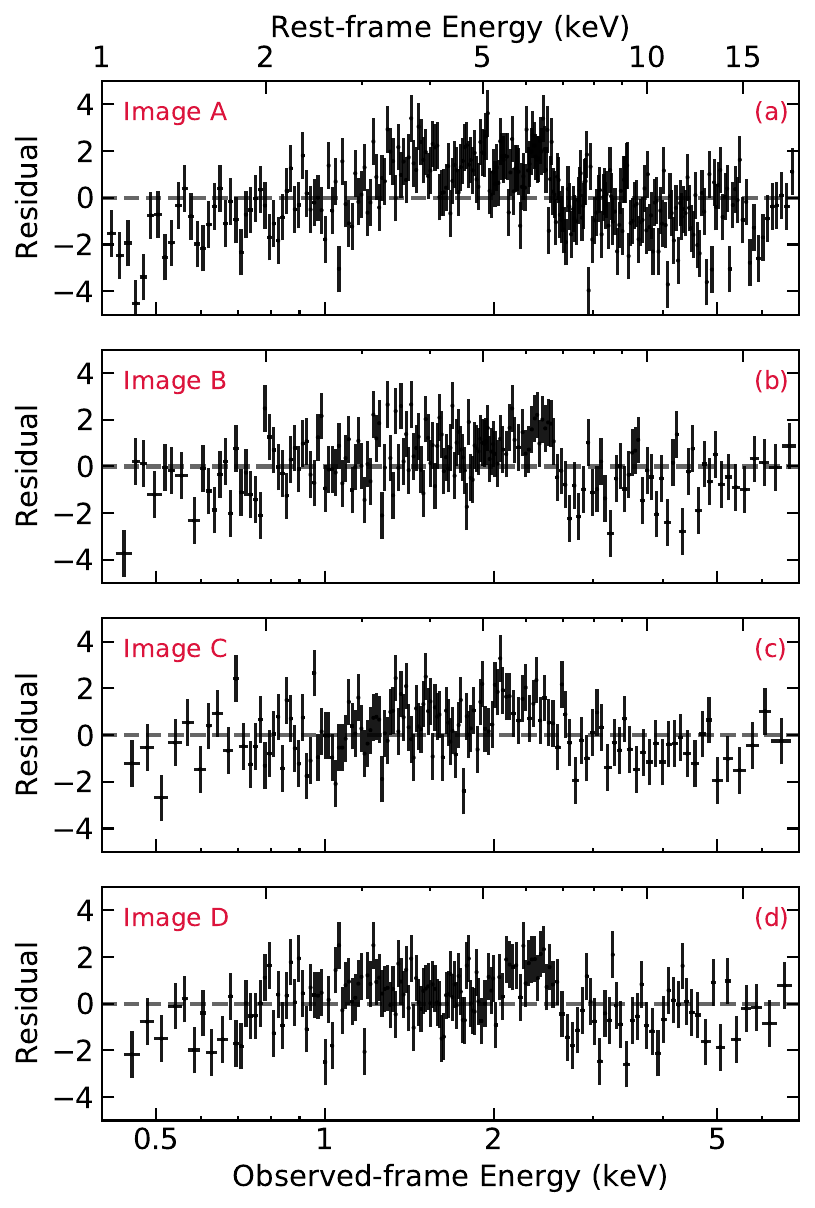}
			\caption{Rest-frame best-fit residuals of the individual-image stacked spectra to Model {\sf pl}. From top to bottom: residuals of image A, B, C and D. }
			\label{fig:cha_stack_single_ima_zpo}
\end{figure}

The analysis applied to these spectra follows the same logic used for the XMM-\textit{Newton} data. In Fig. \ref{fig:cha_stack_single_ima_zpo} we show the best-fit residuals to Model {\sf pl} of the four stacked spectra. The best-fit values of the photon index are consistent among the four images (see Table \ref{tab:cha_stack}). All images show hints of absorption in the soft-energy band but these appear to be more prominent in image A. Moreover, they all present signatures of strong emission lines between 5 and 7 keV. Interestingly, images A, B and D show hints of a line around ${\rm6.4\ keV}$, while for image C it seems to be placed at slightly lower energies (${\rm\simeq5.5\ keV}$). Hints of an absorption line at ${\rm\simeq7.4\ keV}$ seem to be present in images A and D but not in images B and C. Regarding the hard-energy end of the spectra, none show hints of reflection since residuals above 7--8 keV rest frame appear to be quite flat (although noisy). 

We searched for an additional absorption component in each stacked spectrum through the same models used in section \ref{sec:xmm_sample} (Model {\sf pl\_a} and Model {\sf pl\_pca}). All of the four images require extra absorption and are best reproduced by a partial covering medium placed at the quasar's redshift (best-fit values in Table \ref{tab:cha_stack}). By superposing the $\Gamma - N_{\rm H}$ contours, we find that the column density is consistent within 1$\sigma$ errors for all the stacked images. In terms of photon index, ${\rm\Gamma_B}$, ${\rm\Gamma_C}$ and ${\rm\Gamma_D}$ are consistent within 1$\sigma$, while ${\rm\Gamma_A}$, being the steepest, is consistent with the others only within 2.6$\sigma$. Regarding the superposed $N{\rm_H-CF }$ contours, the best-fit values to both the parameters are consistent within 1.6$\sigma$ for all the four images. The best-fit residuals of each spectrum to Model {\sf pl\_pca} still show hints of the lines discussed above. 
\renewcommand{\arraystretch}{1.15}
	\begin{sidewaystable*}
		\caption{Summary of the best-fit parameters of each model tested for the \textit{Chandra} stacked spectra. }
			\centering
			\begin{tabular}{lccccccccc}
			\hline\hline
			Model&Image &${\rm \Gamma}$&$N{\rm_H}$&CF&$E{\rm_{el}}$&${\rm\sigma_{el}}$&$E{\rm_{al}}$&${\rm\sigma_{al}}$& ${\rm\chi^2_r\ (\nu)}$ \\  
			\hline
			{\sf pl}& A   & ${\rm 1.67\pm0.03}$ & -- & -- & -- & -- & -- & -- & 2.20 (260) \\ 
            {\sf pl}& B  & ${\rm 1.64\pm0.03}$ & -- & -- & -- & -- & -- & -- & 1.58 (166) \\ 
            {\sf pl}& C   & ${\rm 1.64\pm0.03}$ & -- & -- & -- & -- & -- & -- & 1.40 (138) \\ 
            {\sf pl}& D   & ${\rm 1.62\pm0.03}$ & -- & -- & -- & -- & -- & -- & 1.39 (153) \\ 
			${\sf pl\_pca}$& A & ${\rm 2.07\pm0.06}$ & ${\rm 4.53_{-0.85}^{+0.88}}$ & ${\rm 0.59\pm0.04}$ & -- & -- & -- & -- & 1.42 (258) \\ 
			${\sf pl\_pca}$& B & ${\rm 1.96\pm0.10}$ & ${\rm 3.44_{-1.49}^{+1.68}}$ & ${\rm 0.55\pm0.08}$ & -- & -- & -- & -- & 1.29 (164) \\ 
			${\sf pl\_pca}$& C & ${\rm 1.98\pm0.14}$ & ${\rm 5.85_{-2.73}^{+2.87}}$ & ${\rm 0.51\pm0.11}$ & -- & -- & -- & -- & 1.22 (136) \\ 
			${\sf pl\_pca}$& D & ${\rm 1.89\pm0.10}$ & ${\rm 2.73_{-1.48}^{+1.70}}$ & ${\rm 0.53\pm0.12}$ & -- & -- & -- & -- & 1.16 (151) \\ 
			${\sf pl\_pca\_el}$& A & ${\rm 2.05\pm0.06}$ & ${\rm 4.09_{-0.82}^{+0.86}}$ & ${\rm 0.58\pm0.04}$           & ${\rm 6.49\pm0.08}$           & ${\rm <0.01}$ & -- & -- & 1.28 (256) \\ 
			${\sf pl\_pca\_el}$& B & ${\rm 1.95\pm0.10}$ & ${\rm 3.04_{-1.42}^{+1.58}}$ & ${\rm 0.55\pm0.08}$           & ${\rm 6.60_{-0.22}^{+0.13}}$  & ${\rm <0.01}$ & -- & -- & 1.20 (162) \\ 
			${\sf pl\_pca\_el}$ (1)& C & ${\rm 1.91\pm0.14}$ & ${\rm 4.45_{-2.97}^{+3.11}}$ & ${\rm 0.46_{-0.14}^{+0.11}}$ & ${\rm 5.56\pm0.08}$ & ${\rm <0.01}$ & -- & -- & 1.12 (134) \\ 
			${\sf pl\_pca\_el}$ (2)& C & ${\rm 1.95\pm0.14}$ & ${\rm 5.08_{-2.77}^{+2.90}}$ & ${\rm 0.49_{-0.13}^{+0.10}}$ & ${\rm 6.37_{-0.34}^{+0.20}}$ & ${\rm <0.01}$ & -- & -- & 1.19 (134) \\ 
			${\sf pl\_pca\_el}$& D & ${\rm 1.89\pm0.10}$ & ${\rm 2.40_{-1.41}^{+1.59}}$ & ${\rm 0.53_{-0.11}^{+0.18}}$  & ${\rm 6.50_{-0.23}^{+0.13}}$  & ${\rm <0.01}$ & -- & -- & 1.07 (149) \\ 
			${\sf pl\_pca\_bel}$& A & ${\rm 2.05\pm0.06}$ & ${\rm 3.93\pm0.84}$ & ${\rm 0.58\pm0.04}$ & ${\rm 6.45\pm0.08}$ & ${\rm 0.16\pm0.08}$ & -- & -- & 1.27 (255) \\ 
			${\sf pl\_pca\_bel}$& B & ${\rm 1.92\pm0.10}$ & ${\rm 2.32_{-1.45}^{+1.58}}$ & ${\rm 0.54_{-0.11}^{+0.19}}$ & ${\rm 6.18_{-0.30}^{+0.43}}$  & ${\rm 0.47_{-0.35}^{+0.23}}$ & -- & -- & 1.17 (161) \\ 
			${\sf pl\_pca\_bel}$ & C & ${\rm 1.85\pm0.13}$ & ${\rm 2.52_{-2.20}^{+3.28}}$ & ${\rm 0.42_{-0.16}}$ & ${\rm 5.94\pm0.23}$ & ${\rm 0.52_{-0.17}^{0.24}}$ & -- & -- & 1.09 (133) \\ 
			${\sf pl\_pca\_bel}$& D & ${\rm 1.88\pm0.10}$ & ${\rm 1.90_{-1.38}^{+1.50}}$ & ${\rm 0.54_{-0.13}}$  & ${\rm 6.28\pm0.20}$  & ${\rm0.41_{-0.18}^{+0.14}}$ & -- & -- & 1.00 (148) \\ 
			${\sf pl\_pca\_el\_al}$& A & ${\rm 2.05\pm0.06}$ & ${\rm 4.18\pm0.85}$ & ${\rm 0.58\pm0.04}$ & ${\rm 6.49\pm0.08}$ & ${\rm <0.01}$ & ${\rm 7.26_{-0.15}^{+0.19}}$ & ${\rm <0.01}$ & 1.27 (254) \\ 
			${\sf pl\_pca\_el\_al}$& B & ${\rm 1.94\pm0.10}$ & ${\rm 3.11_{-1.46}^{+1.63}}$ & ${\rm 0.54\pm0.09}$ & ${\rm 6.60_{-0.22}^{+0.14}}$ & ${\rm <0.01}$ & ${\rm7.49_{-0.21}^{+0.19}}$ & ${\rm <0.01}$ & 1.17 (160) \\ 
			${\sf pl\_pca\_el\_al}$& C & ${\rm 1.91\pm0.14}$ & ${\rm 4.60_{-2.99}^{3.20}}$ & ${\rm 0.45_{-0.14}^{+0.11}}$ & ${\rm 5.57\pm0.08}$ & ${\rm <0.01}$ & ${\rm 7.57_{-0.22}^{+0.40}}$ & ${\rm <0.01}$ & 1.12 (132) \\ 
			${\sf pl\_pca\_el\_al}$& D & ${\rm 1.87\pm0.10}$ & ${\rm 2.43_{-1.44}^{+1.63}}$ & ${\rm 0.53_{-0.11}^{+0.17}}$ & ${\rm 6.50_{-0.23}^{+0.13}}$ & ${\rm <0.01}$ & ${\rm 7.41\pm0.17}$ & ${\rm <0.01}$ & 1.04 (147) \\ \hline
			{\sf pl\_pca\_el}& A+B+D & ${\rm 1.99\pm0.05}$ & ${\rm 3.65\pm0.69}$ & ${\rm 0.54\pm0.04}$ & ${\rm 6.50\pm0.06}$ & ${\rm>0.01}$ & -- & -- & 1.27 (305)\\
			{\sf pl\_pca\_bel}& A+B+D & ${\rm 1.98\pm0.05}$ & ${\rm 3.36\pm0.70}$ & ${\rm 0.54\pm0.04}$ & ${\rm 6.43_{-0.26}^{+0.07}}$ & ${\rm0.23_{-0.08}^{+0.23}}$ & -- & -- & 1.21 (304)\\ \hline
			${\sf pl}$& A+B+C+D & ${\rm 1.66\pm0.02}$ & -- & -- & --  & -- & -- & -- & 2.48 (319) \\ 
			${\sf pl\_pca}$& A+B+C+D & ${\rm 2.00\pm0.04}$ & ${\rm 4.43_{-0.69}^{+0.71}}$ & ${\rm 0.55\pm0.03}$ & -- & -- & -- & -- & 1.53 (317) \\ 
			{\sf pl\_pca\_el}& A+B+C+D & ${\rm 1.99\pm0.04}$ & ${\rm 3.48\pm0.68}$ & ${\rm 0.52\pm0.04}$ & ${\rm 6.49\pm0.07}$ & ${\rm <0.01}$ & -- & -- & 1.32 (315) \\ 
			{\sf pl\_pca\_el\_al}& A+B+C+D & ${\rm 1.98\pm0.04}$ & ${\rm 3.92\pm0.69}$ & ${\rm 0.53\pm0.04}$ & ${\rm 6.49\pm0.07}$ & ${\rm <0.01}$ & ${\rm 7.43\pm0.09}$ & ${\rm <0.01}$ & 1.27 (313)\\ 
			{\sf pl\_pca\_bel}& A+B+C+D & ${\rm 1.95\pm0.04}$ & ${\rm 2.92\pm0.70}$ & ${\rm 0.51\pm0.04}$ & ${\rm 6.09\pm0.14}$ & ${\rm0.52\pm0.11}$ & -- & -- & 1.23 (314)\\
			{\sf pl\_pca\_bel\_al}& A+B+C+D & ${\rm 1.94\pm0.05}$ & ${\rm 2.86\pm0.73}$ & ${\rm 0.50\pm0.04}$ & ${\rm 6.13\pm0.14}$ & ${\rm0.61_{-0.13}^{+0.16}}$ & ${\rm 7.36\pm0.11}$ & ${\rm <0.01}$ & 1.17 (312)\\
			{\sf pl\_pca\_bel\_bal}& A+B+C+D & ${\rm 1.93\pm0.05}$ & ${\rm 2.84\pm0.75}$ & ${\rm 0.50\pm0.04}$ & ${\rm 6.30_{-0.20}^{+0.24}}$ & ${\rm0.73\pm0.16}$ & ${\rm 7.28\pm0.10}$ & ${\rm 0.26_{-0.11}^{+0.16}}$ & 1.15 (311)\\ 
		    {\sf pl\_pca\_bel\_el\_al}& A+B+C+D & ${\rm 1.96\pm0.04}$ & ${\rm 3.26\pm0.68}$ & ${\rm 0.52\pm0.04}$ & ${\rm 6.44\pm0.07}$ & ${\rm0.22\pm0.07}$ & ${\rm 7.42\pm0.10}$ & ${\rm <0.01}$ & 1.14 (310)\\
			  &  &   &   &   &  ${\rm 5.62\pm0.09}$ &   ${\rm <0.01}$ &   &  & \\
			\hline
			\end{tabular}
		\label{tab:cha_stack}
		\tablefoot{Col. 1: model name; Col. 2: combined images; Col. 3: column density in excess to the Galactic value (units of ${\rm 10^{22}\ cm^{-2}}$); Col. 4: covering fraction of the extra absorption; Col. 5: energy of the emission line (in units of keV); Col. 6: width of the emission line (in units of keV); Col. 7: energy of the absorption line (in units of keV); Col. 6: width of the absorption line (in units of keV); Col. 7: Reduced ${\rm\chi^2}$ (degrees of freedom). All the errors are evaluated at 90\% confidence. \textit{Model list}: Model {\sf pl}={\tt phabs*zpo}; Model {\sf pl\_pca}={\tt phabs*zpcf*zpo}; Model {\sf pl\_pca\_el}={\tt phabs*zpcf*(zpo+zgauss)}; Model {\sf pl\_pca\_el\_al}={\tt phabs*zpcf*(zpo+zgauss+zgauss)}. When the width of the emission/absorption line is a free parameter of the model, it is reported as a bel/bal component in its name. }
	\end{sidewaystable*}
\renewcommand{\arraystretch}{1}

Images A, B and D show a highly significant Fe K$\alpha$ emission line between $\simeq6.4$--6.6 keV, that we first constrained as a narrow Gaussian component (Model {\sf pl\_pca\_el}, Table \ref{tab:cha_stack}). Regarding image C, we find a highly significant narrow line placed at lower energies (${\rm\simeq5.6\ keV}$), while we only marginally detect the Fe K$\alpha$ at ${\rm6.4\ keV}$ (see Table \ref{tab:cha_stack}). Given the results obtained by \citet{reynolds2014} over the combined spectrum of all images from all epochs, we searched for broad emission lines in the single-image stacked spectra. When letting the width of these lines vary, the improvement in the goodness of the fit is highly significant only for image D (${\rm\Delta\chi^2=11.0}$ for ${\rm\Delta\nu=1}$), with new best-fit parameters $E_{\rm rf}=6.28\pm0.20\ {\rm keV}$ and ${\rm\sigma=0.41_{-0.18}^{+0.14}\ keV}$. We marginally detect a broad emission line in image A (${\rm E=6.45\pm0.08\ keV}$, ${\rm\sigma=0.16\pm0.09\ keV}$) and image B ($E_{\rm rf}=6.18_{-0.30}^{+0.43}\ {\rm keV}$, ${\rm\sigma=0.47_{-0.35}^{+0.23}\ keV}$), while for image C the data are best reproduced by a narrow line. The resolved width found for the emission lines can be explained as follows. From the HSS (see Sect. \ref{sec:cha_lines}), we know that the Fe K$\alpha$ is probably microlensed, thus its energy likely varies from epoch to epoch at fixed image. 
Detecting a broad Fe K$\alpha$ line in the single-image stacked spectra can be interpreted as indicating that the microlensing of the line at energies near the intrinsic energy of ${\rm6.4\ keV}$ is the most frequent effect, while those events producing more extreme energy shifts are more rare ore less effective, as also shown in \citet{chartas2016_gravlens} for RX J1131-1231. 

The narrow absorption line is marginally detected (at 90\%-99\% confidence) at $E_{\rm rf}\simeq7.3$--$7.5\ {\rm keV}$ in the stacked spectra of image A, B and D (best-fit parameters in Table \ref{tab:cha_stack}). However, as the residuals in Fig. \ref{fig:cha_stack_single_ima_zpo} suggest, it is unrequired for image C. 

To summarize, we find that the cold and partial covering absorber and intrinsic emission are common to all the individual-image stacked spectra, although image A shows a slightly steeper photon index. This supports our previous result of the absorber's column density being dominated by the in situ component (see Section \ref{sec:cha_sample}). Regarding the narrow features, however, it is harder to provide a uniform description of the four spectra. On the one hand, image C is surely the most peculiar since it is the only one that presents an emission line at ${\rm\approx5.6\ keV}$ (consistent with being narrow) and a marginally detected Fe K$\alpha$. In addition, it does not even show hints of the absorption line at ${\rm\simeq7.4\ keV}$. On the other hand, the energies of the narrow Fe K$\alpha$ from all the stacked images (also that of the marginal detection in image C) are consistent within 1$\sigma$, and so they remain if we consider the energy of the broad lines. 

Moreover, being the stacking of 37 epochs, we should bear in mind that these broad emission lines might not be intrinsically broad. They could likely be produced by the combination of single-epoch microlensed Fe K$\alpha$, which we actually see in the single-image spectra (Sect. \ref{sec:cha_lines}). In this sense, the differences in significance we find between the four images are to be interpreted as the result of different microlensing events occurring in the respective image through the epochs. 

In conclusion, when stacking all the single-image spectra from all the epochs into one single spectrum, we should take all the properties found above into account, especially those we find in image C. 

\subsection{Stacked spectra of all images}
\label{app:all_combined}
\begin{figure}[!h]
			\centering
			\includegraphics[width=1.0\linewidth]{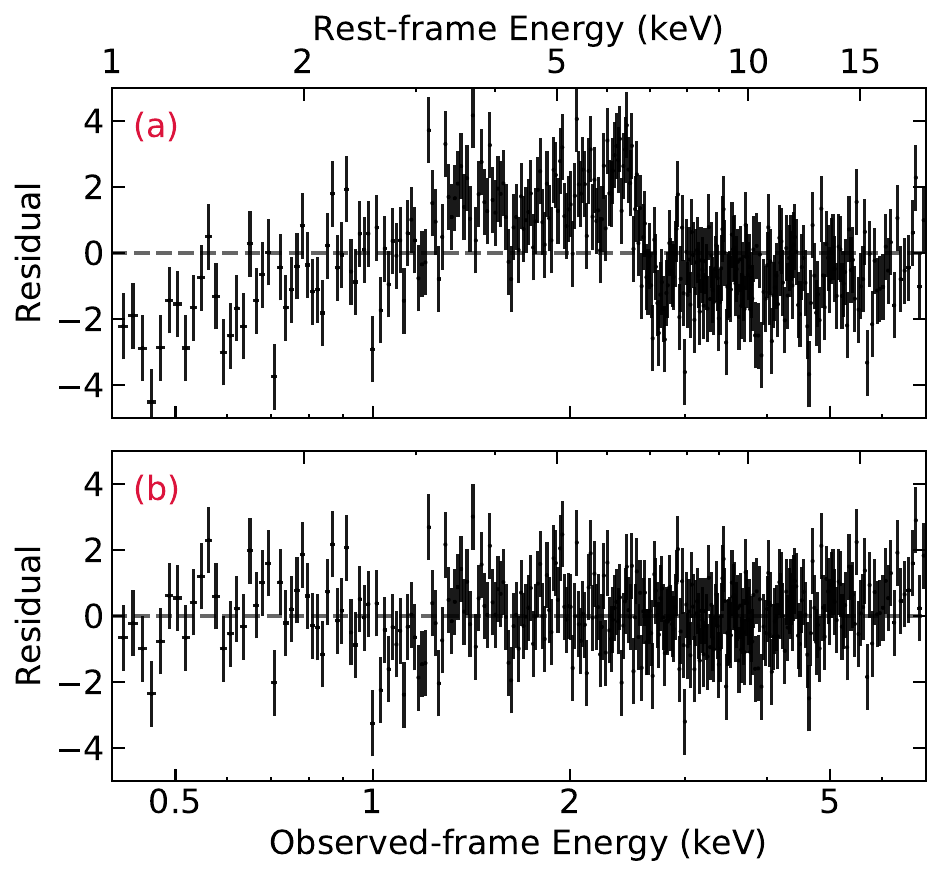}
			\caption{Best-fit residuals of the final \textit{Chandra} spectrum to Model {\sf pl} (\textit{panel a}) and Model {pl\_pca\_bel\_el\_al} (\textit{panel b}). }
			\label{fig:cha_stack_all}
\end{figure}

The final stacked spectrum sums up to a total exposure time of almost 3Ms and 32032 source net counts in the 0.4--7 keV observed-energy range. Its best-fit residuals to Model {\sf pl} are shown in Fig. \ref{fig:cha_stack_all}, panel a. As expected from the results in the previous section, we find evidence of absorption in the soft-energy band, a prominent emission line at ${\rm\approx6.5\ keV}$ that seems skewed at lower energies and hints of an absorption line at ${\rm\approx7.4 \ keV}$. 
Adding a partial covering cold absorber significantly improves the quality of the fit (see Table \ref{tab:cha_stack}). The best-fit parameters we obtain agree with those of the single images and the residuals still show an excess at ${\rm\approx6.5\ keV}$ and a deficit of counts at ${\rm\approx7.4\ keV}$. By adding a narrow Gaussian component each, we find a highly significant emission line at $E_{\rm rf}=6.49\pm0.07\ {\rm keV}$ (${\rm\Delta\chi^2/\Delta\nu=70/2}$) and an absorption line at $E_{\rm rf}=7.43\pm0.09\ {\rm keV}$ (${\rm\Delta\chi^2/\Delta\nu=18/2}$). However, the residuals still indicate the emission line to be skewed at lower energies, so we let its width vary (Model {\sf pl\_pca\_bel\_al}). We find a much better best-fit (${\rm\Delta\chi^2=32.3}$, ${\rm\Delta\nu=1}$), with a broad emission line at $E_{\rm rf}=6.13\pm0.14\ {\rm keV}$ and ${\rm\sigma=0.61_{-0.13}^{+0.16}\ keV}$. When allowing also the absorption-line width to vary (Model {\sf pl\_pca\_bel\_bal}), we find it significantly (${\rm\Delta\chi^2=10.1}$, ${\rm\Delta\nu=1}$) consistent with being broad (${\rm\sigma=0.26_{-0.11}^{+0.16}\ keV}$) and placed at $E_{\rm rf}=7.28\pm0.10\ {\rm keV}$. This changes also the centroid and the width of the emission line, which become $E_{\rm rf}=6.29_{-0.20}^{+0.24}\ {\rm keV}$ and ${\rm\sigma=0.72\pm0.16\ keV}$. However, the best-fit residuals of Model {\sf pl\_pca\_bel\_bal} indicate that the model is overestimating/underestimating the data at energies lower/higher than the best-fit centroid of the emission line. This could be the indication of a relativistically blurred Fe K$\alpha$ line. In fact, the energy and width we find using Model {\sf pl\_pca\_bel\_al} are consistent with those found by \citet{reynolds2014}, which they interpret as the indication of a relativistically broadened line. If we exclude the absorption line from the model, we find the same skewed centroid energy and width as in the preliminary analysis done by \citet{reynolds2014}. 

However, based on the single-image stacked-spectra results, the width of the emission line could be artificially produced by the stacking, since the line in image C is placed at lower energies w.r.t the other three (see Fig. \ref{fig:cha_stack_single_ima_zpo}). Thus, we produced a new stacked spectrum, combining all the epochs of only images A, B and D. Considering the absorber and the emission feature (Model {\sf pl\_pca\_bel}), we find that the line is less broad (${\rm\sigma=0.23_{-0.08}^{+0.23}\ keV}$) and, more importantly, its centroid is placed at higher energies ($E_{\rm rf}=6.43_{-0.26}^{+0.07}\ {\rm keV}$). Moreover, constraining the width of the line to be lower than the CCD resolution or letting it to vary freely make almost no difference on a statistical basis (see Table \ref{tab:cha_stack}). Finally, if we compute the F-test significance for the addition of the width as a free parameter, we find that it is not required by the data. Thus, the skewed emission line we find at ${\rm\approx6.13\ keV}$ is most probably generated by the blending of two distinct lines. 

Given the results obtained with the combined spectrum of images A+B+D and those from the individual-image stacked spectra, we tried to model the skewed emission line as a narrow component plus a broad component, the first at ${\rm\approx 5.6\ keV}$ and the second at ${\rm\approx 6.5\ keV}$ (Model {\sf pl\_pca\_bel\_el}). This produced a ${\rm\Delta\chi^2=17.1}$ for two parameters of interest (w.r.t Model {\sf pl\_pca\_bel}), that according to the F-test translates in a detection above ${\rm99.99\%}$ confidence of the narrow line placed at $E_{\rm rf}=5.62\pm0.08\ {\rm keV}$. The broad Fe K$\alpha$ is now detected at $E_{\rm rf}=6.44\pm0.07\ {\rm keV}$ with ${\rm \sigma=0.22\pm0.07\ keV}$, which is inconsistent to the centroid energy of the relativistically skewed line found by \citet{reynolds2014} ($E_{\rm rf}=6.58\pm0.03\ {\rm keV}$). We also detect the narrow absorption line at $E_{\rm rf}=7.42\pm0.10\ {\rm keV}$ (Model {\sf pl\_pca\_bel\_el\_al}) at 99.8\% confidence (from the F-test). This model also corresponds to the one that returns the best representation of the data, both on the basis of the distribution of the residuals (see Fig. \ref{fig:cha_stack_all}, panel b) and in terms of statistical improvement. This result corroborates our statement of the skewed line being the blending of the two lines we find in the individual-image staked spectra. Thus, when stacking the spectra of a gravitationally lensed quasar, checking the properties of each image is fundamental. 

In conclusion, the spectral features of the stacked spectra confirm the presence of two distinct outflow components based on the following arguments. 
The absorption lines at higher energies (${\rm\approx11.8\ keV}$) of the single-epoch \textit{Chandra} spectra are absent in the stacked spectra, whereas we detect (marginally or significantly) the absorption line at ${\rm\approx7.4\ keV}$. The ${\rm11.8\ keV}$ features are probably associated to an outflow whose ionization state (i.e. the absorption line energy) varies more rapidly than that of the wind producing the ${\rm7.4\ keV}$ line. Thus, it indicates that the two winds are consistent with being launched at different radii w.r.t. the central engine, i.e. the one associated with the ${\rm11.8\ keV}$ being produced closer to the BH. Moreover, this scenario would also agree with that proposed in Sect. \ref{sec:disc_res} based on the significance of the two lines in XMM 2002. This would imply the absence of lines in XMM 2018 as the indication that during the 2018 observation either both outflows are weak (in terms of velocity component along the line of sight) or that the outermost is weak and the other is so extremely ionized that it becomes undetectable. 

\end{appendix}

\end{document}